\documentclass[preprint,12pt,sort&compress]{elsarticle}

\usepackage{amssymb}
\usepackage{epsfig}
\usepackage{hyperref}

\journal{Nuclear Physics A}

\begin{document}

\begin{frontmatter}



\title{Constraining mean-field models of the nuclear matter equation
of state at low densities}


\author{ M.D. Voskresenskaya}
\author{S. Typel}

\address{GSI Helmholtzzentrum f\"{u}r Schwerionenforschung GmbH,
Theorie, Planckstra\ss{}e 1, D-64291 Darmstadt, Germany}

\begin{abstract}
An extension of the generalized relativistic mean-field (gRMF)
model with density dependent couplings is introduced in order
to describe thermodynamical properties and the composition of
dense nuclear matter for astrophysical applications.
Bound states of light nuclei and two-nucleon scattering correlations
are considered as explicit degrees of freedom in the thermodynamical
potential. They are represented by quasiparticles with medium-dependent properties.
The model describes the correct low-density limit given by the virial equation of state (VEoS) 
and reproduces RMF results around nuclear saturation density where clusters are dissolved.
A comparison
between the fugacity expansions of the VEoS and the gRMF model
provides consistency relations between the quasiparticles properties,
the nucleon-nucleon scattering phase shifts and
the meson-nucleon couplings of the gRMF model at zero
density. Relativistic effects are found to be important at
temperatures that are typical
in astrophysical applications. Neutron matter and symmetric
nuclear matter are studied in detail.
\end{abstract}


\end{frontmatter}


\newpage


\section{Introduction}

Correlations are an essential feature of an interacting many-body
system such as nuclear matter. They have an impact on the thermodynamical
properties and the composition of the matter since the constituent particles
can form new nuclear species as bound states. 
This aspect is particularly important in the description of matter below nuclear
saturation density that is employed in theoretical simulations of
astrophysical phenomena
such as core-collapse supernovae. There, the knowledge of the equation
of state (EoS) of neutron-proton asymmetric matter is required
in a wide range of densities and temperatures \cite{Kla06,Jan07}.
The description of dense matter is important
for the investigation of various stages in supernova explosions and
the structure of neutron stars \cite{Gle00}.
It is known to affect the effectiveness of the neutrino reheating
of the shock wave \cite{Lat04}.
The sensitivity of the collapse dynamics on the properties of
matter and the composition in this density regime could strongly influence the
structure of the proto-neutron star \cite{Sum08}.

Although there are many
approaches to describe dense matter, in particular at zero
temperature, only few models cover the full
parameter space needed in most astrophysical applications. Often, they
do not supply sufficient information on the thermodynamical and compositional
details. Thus, for many years, a very small number of EoS tables was
available which have been
used in simulations of dynamical astrophysical processes
\cite{Lat91,HShe98a,HShe98b}.
However, in
recent years, the interest in developing EoS has surged and several
new EoS tables were provided
\cite{Ish08,Ohn10,Hem10,GShe10a,GShe10b,GShe11a,GShe11b,HShe11}.
This development was triggered by the needs of astrophysical modelers,
the progress in the theoretical description of nuclear matter and
an increase of computational power. Nevertheless,
approximations and simplifications are still needed for practical
purposes.

There are two major paths to build an EoS of warm and dense nuclear
matter for practical applications. One approach starts with an
ideal mixture of nucleons and nuclei leading to a nuclear statistical
equilibrium (NSE) description \cite{Hem10}. The effect of interactions between
all constituents can be incorporated with the help of virial
corrections. However, at present, only nucleons and light nuclei are
considered in practice in such a virial equation of state (VEoS)
\cite{Hor06a,Hor06b,OCo07} that
provides the correct finite-temperature EoS in the limit of very
low densities. The results are model independent since they depend
only on experimentally determined data, i.e.\ binding energies of
nuclei and scattering phase shifts.
Unfortunately, the application of this approach is limited to rather low
densities. The dissolution of nuclei and the
transition to uniform neutron-proton matter with increasing density cannot be
described properly. In order to simulate such an effect, the heuristic
excluded volume mechanism was employed frequently \cite{Hem10}. A second class of
EoS models for astrophysical applications
is based on self-consistent mean-field methods with neutrons and
protons as fundamental constituents. They are considered as
quasiparticles with self-energies that contain the information on
the interaction which is usually modeled in an effective way and not
taken from a realistic nucleon-nucleon (NN) interaction. These mean-field models,
both non-relativistic and relativistic,
can be very successful in describing finite nuclei
and nuclear matter around saturation density
$n_{\rm sat} \approx 0.16$~fm \cite{Gam90,Ser96,Ben03,Kol05}.
But at low densities $n \ll n_{\rm sat}$ they fail to
include correlations properly
that give rise to inhomogeneities and the formation of many-body bound states,
i.e.\ nuclei. There were attempts to modify the low-density behavior
of mean-field models guided by microscopic calculations, see,
e.g., Ref.\ \cite{Mar07} for the zero temperature case. However,
it is still challenging to include the formation
of clusters properly.
The deficiencies of the various models
lead to the strategy of patching up or merging different approaches providing
a uniform description of warm dense matter.

A recent publication \cite{Typ10}
considered a quantum statistical approach to
nuclear matter and devised
a generalized relativistic mean-field (gRMF) model, which is an extension of the
relativistic mean-field model  with density dependent couplings
(DD-RMF) from 
Ref.\ \cite{Typ99}.
The gRMF model includes, besides nucleons,
light nuclei (${}^{2}$H$\: =d$, ${}^{3}$H$\: =t$, ${}^{3}$He$\: =h$,
${}^{4}$He$\: =\alpha$) as hadronic degrees of freedom in the Lagrangian.
The dissolution of these clusters is modeled by a medium-dependent
shift of their binding energies originating mainly from the action
of the Pauli principle. All hadronic constituents of the model
can be  viewed as quasiparticles that
interact via the exchange of effective mesons.
The gRMF model achieves bridging
the two views on nuclear matter, i.e.\ a mixture of nucleon and nuclei at low
densities and nucleons as quasiparticles moving
in mean fields at high densities.
Even though the model contains the
same relevant particles at low densities as the VEoS, it does not
reproduce exactly the thermodynamical properties of the VEoS as will
be shown below.
In fact, all presently employed EoS for astrophysical applications
that are based on mean-field concepts
share the problem of reproducing the VEoS in the low-density limit.
Thus the question arises, how the models
can be modified in order to approach the correct low-density
limit. Is a modification of the effective interaction sufficient to meet this
aim or are there more severe changes of the models required?

In this work we propose an approach
that introduces bound and scattering states in
an effective way as additional degrees of freedom
in the thermodynamical potential. All relevant quantities are derived
in a thermodynamically consistent way. These
many-body bound and scattering states are represented by 
quasiparticles with medium-dependent properties
and temperature dependent energies.
The low-density behavior of nuclear matter at finite temperatures
is considered in detail by comparing the VEoS
with the gRMF approach by means of a series expansion of the
grand
canonical
potential 
in powers of the nucleon fugacities.
From the comparison of the VEoS and gRMF expansions we
 will derive consistency relations that connect quasiparticle 
properties with the meson-nucleon
couplings in the vacuum and the phase shifts or
effective-range parameters of nucleon-nucleon scattering.
We investigate different choices of the meson-nucleon
couplings and quasiparticle properties set by these consistency relations.
In addition, relativistic corrections to the traditional
VEoS are obtained from these relations that become larger than
 the effects of particle correlations in the low-density limit.
Various parametrizations of the quasiparticle
medium dependence are discussed.
Studying  thermodynamical properties of the
matter  we consider temperatures higher
than the critical temperatures for pairing and Bose condensation
in order to avoid these more complex  effects. Thus our consideration of the
low and zero temperature limits should be corrected in the 
future to incorporate condensation phenomena.
Our approach is not limited to the
chosen particular gRMF model with density-dependent meson-nucleon
couplings. It can be applied easily to other mean-field
approaches as well.

This paper is organized as follows. In Section \ref{sec:veos} we present
a general formulation of the VEoS that serves as one starting point of the
discussion. Basic quantities are defined and the notation is
established. We show that correlations in continuum states can be represented
effectively by temperature dependent resonances.
Analytic expressions for the second virial coefficients
are found with the help of the effective-range expansion for the
s-wave phase shifts. The connection of the VEoS and NSE models is established.
The relation between the VEoS and the generalized Beth-Uhlenbeck approach, 
which takes into account modifications due to medium effects,
is discussed.
The gRMF model with
density-dependent meson-nucleon couplings is sketched in Section
\ref{sec:gRMF} and a series expansion of the
grand canonical
potential $\Omega$ in powers of fugacities is derived for low densities.
The power series for $\Omega$ in the VEoS and gRMF are compared
up to second-order
leading to consistency
relations between the models that are studied in various limits.  
In Sect. \ref{sec:ExtRMF}
we consider an extension of the gRMF model:
in order to satisfy
consistency relations we introduce temperature 
dependent resonance energies of the scattering states and modified degeneracy factors
that allow to make a smooth interpolation between VEoS and gRMF models.
We find that relativistic corrections to the standard VEoS
are important in these consistency relations  already in the first order of the
expansion.
The particular example of neutron matter is presented in
Section \ref{sec:neumat} and symmetric nuclear matter is discussed
in Section \ref{sec:snm}.
The following section deals with the transition from low to high
densities and the occurring problems.
Concluding remarks and an outlook are given in Section
\ref{sec:concl}. The comparison of our notations 
with the ones given in Ref.\ \cite{Hor06b}
can be found in \ref{sec:HS}.
An expansion of the energy per particle in neutron
matter at zero temperature in powers of the Fermi momentum is
considered in \ref{sec:RMFlow}.
Throughout this work we use natural units such that
$\hbar = c = 1$.

\section{Equation of state in the virial limit}
\label{sec:veos}

\subsection{General formalism}

The VEoS represents a model-independent approach
in the calculation of thermodynamical properties of
low-density matter at finite temperatures $T$.
In the non-relativistic limit, which is usually considered,
it describes a system of interacting
particles $i,j,\dots$ with non-relativistic chemical 
potentials $\mu_{i}$ in a volume $V$,
provided the fugacities $z_{i}=\exp (\mu_{i}/T)$
 are small ($z_{i}\ll1$).
The range of densities where
the virial expansion is valid can be estimated by the relation
$ n_{i}\lambda_{i}^{3} \ll 1$
where $n_{i}$ is the number density
of particle $i$ with mass $m_{i}$ and $\lambda_{i}=\sqrt{2\pi/(m_{i} T)}$
is the thermal wavelength.

The following presentation gives a generalized and more symmetric
formulation of the
virial approach as compared to \cite{Hor06a,Hor06b,OCo07}. The S-matrix
formulation by Ref.\ \cite{Das69}
that was recently applied to nuclear matter in Ref.\ \cite{Mal08} 
gives essentially identical results.

Under the presupposed conditions the
grand canonical partition function $\mathcal{Q}$
can be expanded in powers of fugacities as
\begin{equation}
\label{eq:Qdef}
 \mathcal{Q}(T,V,\mu_{i}) =
 1+\sum_{i} Q_{i}z_{i}
 +\frac{1}{2}\sum_{ij} Q_{ij}z_{i}z_{j}
 +\frac{1}{6}\sum_{ijk} Q_{ijk} z_{i}z_{j}z_{k} + \dots
\end{equation}
with one-, two-, three-, \dots many-body canonical
partition functions $Q_i$, $Q_{ij}$, $Q_{ijk}$, \dots.
In classical non-relativistic mechanics, the single-particle
canonical partition function is
\begin{equation}
 Q_{i}  =
 \frac{g_{i}}{(2\pi)^3} \int d^{3}r_{i} \int d^{3}p_{i}
 \: \exp\left(-\beta H_{i}\right) = \frac{g_{i}}{\lambda_{i}^{3}} V
\end{equation}
with the spin degeneracy factor $g_{i}$ ($=2$ for each nucleon),
$\beta = 1/T$
and the single-particle Hamiltonian $H_{i} = p_{i}^{2}/(2m_{i})$
that only contains a kinetic contribution with momentum $p_{i}$.
For the two-body canonical partition function we have
\begin{equation}
 Q_{ij} = \frac{g_{i} g_{j}}{(2\pi)^{6}}
 \int d^{3}r_{i}\int d^{3}p_{i} \int d^{3}r_{j} \int d^{3} p_{j}
 \: \exp\left(-\beta H_{ij}\right)
\end{equation}
with the two-body Hamiltonian
$H_{ij} = p_{i}^{2}/(2m_{i}) + p_{j}^{2}/(2m_{j}) + V_{ij}$
that includes a two-body potential $V_{ij}$ responsible for the
correlations.

Then, the grand canonical potential
\begin{equation}
 \Omega(T,V,\mu_{i}) = -T\ln \mathcal{Q}(T,V,\mu_{i})=-pV \: ,
\end{equation}
that is directly related to the pressure $p$,
can be written in the form
\begin{eqnarray}
\label{eq:Omega_series}
 \lefteqn{\Omega(T,V,\mu_{i}) =}
 \\ \nonumber & &  -T V \left(
 \sum_{i} \frac{b_{i}}{\lambda_{i}^{3}} z_{i}
 +  \sum_{ij} \frac{b_{ij}}{\lambda_{i}^{3/2}\lambda_{j}^{3/2}} z_{i}z_{j}
 +  \sum_{ijk} \frac{b_{ijk}}{\lambda_{i}\lambda_{j}\lambda_{k}}
 z_{i}z_{j}z_{k} + \dots  \right)
\end{eqnarray}
with the (symmetrized) dimensionless cluster (or virial) coefficients
\begin{eqnarray}
\label{eq:bi}
 b_{i}(T) & = & \frac{\lambda_{i}^{3}}{V} Q_{i} \: ,
 \\
 b_{ij}(T) & =& \frac{\lambda_{i}^{3/2}\lambda_{j}^{3/2}}{2V}\left(
 Q_{ij} - Q_{i}Q_{j} \right) \: ,
 \\
 b_{ijk}(T) & = & \frac{\lambda_{i} \lambda_{j} \lambda_{k}}{6V}
 \left(Q_{ijk}-Q_{i}Q_{jk}-Q_{j} Q_{ik}
 -Q_{k} Q_{ij}+2Q_{i} Q_{j} Q_{k}\right)
\end{eqnarray}
that depend on the temperature.
Without interaction, the two-, three-, \dots many-body partition functions
factorize, e.g.\ $Q_{ij} = Q_{i}Q_{j}$, and the second, third, \dots cluster
coefficients $b_{ij}$, $b_{ijk}$, \dots vanish.

Individual particle number densities are found from the relation
\begin{equation}
\label{eq:ni_veos}
 n_{i} = - \frac{1}{V} \left. \frac{\partial \Omega}{\partial \mu_{i}}
  \right|_{T,V,\mu_{j\neq i}} =  b_i
\frac{z_{i}}{\lambda_{i}^{3}}
 +2\sum_{j} b_{ij}\frac{z_{i}z_{j}}{\lambda_{i}^{3/2}\lambda_{j}^{3/2}}
 +3\sum_{jk}b_{ijk} \frac{ z_{i}z_{j}z_{k}}{\lambda_{i} \lambda_{j}
   \lambda_{k}} + \dots
\end{equation}
with contributions from free particles (first term in the second
equation) 
and correlated pairs, triples, etc. The entropy is
\begin{eqnarray}
 S & = & - \left. \frac{\partial \Omega}{\partial T}
  \right|_{V,\mu_{i}}
 \\ \nonumber & = &
 -\frac{5\Omega}{2T}
 + V \left(
 \sum_{i} \frac{c_{i}}{\lambda_{i}^{3}} z_{i}
 +  \sum_{ij}
 \frac{c_{ij}}{\lambda_{i}^{3/2}
   \lambda_{j}^{3/2}} z_{i}z_{j}
 +  \sum_{ijk}
 \frac{c_{ijk}}{\lambda_{i}\lambda_{j}\lambda_{k}}
 z_{i}z_{j}z_{k} + \dots  \right)
\end{eqnarray}
with coefficients
\begin{eqnarray}
 c_{i} & = & T \frac{db_{i}}{dT} -\frac{\mu_{i}}{T} \: ,
 \\
 c_{ij} & = & T \frac{db_{ij}}{dT} -\frac{\mu_{i}+\mu_{j}}{T} \: ,
 \\
 c_{ijk} & = & T \frac{db_{ijk}}{dT}
 -\frac{\mu_{i}+\mu_{j}+\mu_{k}}{T} \: .
\end{eqnarray}
Other relevant thermodynamical quantities such as the free energy
and the internal energy
\begin{equation}
 F=\Omega+V\sum_{i}\mu_{i}n_{i}
\end{equation}
and the internal energy
\begin{equation}
 E=F+TS
\end{equation}
can be obtained immediately.

The formulas given above can be
generalized to include four-, five-, \dots many-body correlations,
but already contributions of the three-body term
are hardly ever considered in practice.
For a given temperature,
two-body correlations will always dominate higher-order
correlations when the density decreases. Hence, in the following,
we will truncate the
expansion at second order in the fugacities of the basic constituents.

The quantum mechanical generalization of the virial expansion
up to second order
was given by Beth and Uhlenbeck \cite{Bet36,Bet37}.
In classical mechanics the interaction potential between the particles
is the relevant quantity that appears in the calculation of the virial
coefficients. In quantum mechanics, the Schr\"{o}dinger equation has to
be solved with this potential and the obtained eigenstates comprise bound
and scattering states.
Integrations over phase space are replaced
by sums over all eigenstates of the one- and two-body system and it is
obvious that the density of states becomes the relevant quantity.
The result for the first virial coefficient $b_{i}$ is identical to
the classical value (\ref{eq:bi}) in the continuum approximation, 
corresponding to the replacement of
the sum over discrete momentum states by an integral. The
second virial coefficient can be expressed as an integral
over center-of-mass energies $E$
\begin{equation}
\label{eq:vircoeff}
 b_{ij}(T)=\frac{1+\delta_{ij}}{2}
 \frac{\lambda_{i}^{3/2}\lambda_{j}^{3/2}}{\lambda_{ij}^{3}}
 \int dE \: \exp\left(-\beta E \right) D_{ij}(E)
 \pm \delta_{ij} g_i  2^{-5/2}
\end{equation}
with $\lambda_{ij} = \sqrt{2\pi/[(m_{i}+m_{j})T]}$ and the quantity
\begin{equation}
\label{eq:Dij}
 D_{ij} (E) = \sum_{k}g_{k}^{(ij)}\delta(E-E_{k}^{(ij)})
  +\sum_{l}\frac{g_{l}^{(ij)}}{\pi}\frac{d\delta_{l}^{(ij)}}{dE}
\end{equation}
that measures the
difference between the density of states of an interacting and
free two-particle system. The last term in equation
(\ref{eq:vircoeff})
is a quantum statistical correction with the positive (negative) sign
if $i=j$ are identical bosons (fermions).

The first contribution to $D_{ij}$ is a sum
over all two-body bound states $k$ of the system $ij$ with degeneracy
factors $g_{k}^{(ij)}$ and energies $E_{k}^{(ij)}<0$. The second term
takes into account continuum correlations in the two-particle system in all
channels $l$ with degeneracy factors $g_{l}^{(ij)}$
via the energy-dependent scattering phase shifts $\delta_{l}^{(ij)}$.
In general, the index $l$ stands for the total and angular orbital
momenta of a particular channel. 
If the phase shift is dominated by a narrow
resonance, it 
jumps by $\pi$ in a very small energy interval around the resonance
energy and the contribution to the virial coefficient resembles that
of a bound state. 
With experimentally known bound state energies and phase shifts,
the second virial coefficient $b_{ij}$ and thus the low-density
EoS can be calculated in a model-independent way.

The quantum mechanical generalization of the third virial coefficient
was discussed in Ref.\ \cite{Pai59} but actual calculations turn out to be difficult
in practice and, thus, they are not considered here.
The derivation of the classical VEoS is based on non-relativistic kinematics.
In order to include relativistic effects, at least the
relativistic dispersion relation of the particles has to be
used. The resulting modifications of the virial coefficients
will be discussed when the VEoS is compared to the gRMF approach
in Section \ref{sec:comp}.

\subsection{Application to nuclear matter with arbitrary neutron to proton ratio}

In low-density nuclear matter at not too high temperatures, 
neutrons and protons are the relevant
fundamental constituents. Hence, the grand canonical potential up 
to second order in the fugacities
becomes
\begin{eqnarray}
\label{eq:Omega_veos}
 \lefteqn{\Omega(T,V,\mu_{n},\mu_{p})}
 \\ \nonumber & = & -T V \left(
 \frac{b_{n}}{\lambda_{n}^{3}} z_{n}
 + \frac{b_{p}}{\lambda_{p}^{3}} z_{p}
 +  \frac{b_{nn}}{\lambda_{n}^{3}} z_{n}^{2}
 +  \frac{b_{pp}}{\lambda_{p}^{3}} z_{p}^{2}
 +  2 \frac{b_{np}}{\lambda_{n}^{3/2}\lambda_{p}^{3/2}} z_{n}z_{p}
 \right)
\end{eqnarray}
where the symmetry $b_{np} = b_{pn}$ is used.
The Coulomb interaction is usually neglected in
the VEoS and $b_{pp}$ is replaced by
$b_{nn}$ in (\ref{eq:Omega_veos}), cf.\ Ref.\ \cite{Hor06b}.
The second virial coefficient $b_{nn}$ receives
contributions only from scattering states. It is convenient to split
$b_{np}=b_{np0}+b_{np1}$ into two contributions with total
isospin $\mathcal{T}=0$ and $\mathcal{T}=1$. The only bound two-body state in the
two-nucleon system, the deuteron, appears in the isospin zero
channel with $g_{d}=3$ and $E^{(np)}_{d} = -B_{d} = -2.225$~MeV \cite{Aud03}.
According to this observation we can write
\begin{eqnarray}
\label{eq:bnn}
 b_{nn}(T) & = & \frac{\lambda_{n}^{3}}{\lambda_{nn}^{3}}
 \sum_{l} g_{l}^{(nn)} I_{l}^{(nn)} - g_{n} 2^{-5/2} \: ,
 \\
\label{eq:bpp}
 b_{pp}(T) & = & \frac{\lambda_{p}^{3}}{\lambda_{pp}^{3}}
 \sum_{l} g_{l}^{(pp)} I_{l}^{(pp)} - g_{p} 2^{-5/2} \: ,
 \\
 b_{np1}(T) & = & \frac{1}{2}
 \frac{\lambda_{n}^{3/2}\lambda_{p}^{3/2}}{\lambda_{np}^{3}}
 \sum_{l} g_{l}^{(np1)} I_{l}^{(np1)} \: ,
 \\
\label{eq:bnp0}
 b_{np0}(T) & = &
 \frac{1}{2}
 \frac{\lambda_{n}^{3/2}\lambda_{p}^{3/2}}{\lambda_{np}^{3}}
 \left[  g_{d} \exp\left( \frac{B_{d}}{T} \right)
 +  \sum_{l} g_{l}^{(np0)} I_{l}^{(np0)} \right]
\end{eqnarray}
with the virial integral
\begin{equation}
\label{eq:virint}
 I_{l}^{(ij)}(T) = \int_{0}^{\infty} \frac{dE}{\pi} \:
 \frac{d\delta_{l}^{(ij)}}{dE} \exp\left(-\frac{E}{T} \right) \: .
\end{equation}
Formally, we can write the sum of the scattering contributions in
Eqs. \ (\ref{eq:bnn})-(\ref{eq:bnp0})
in the form of a single bound state contribution
\begin{equation}
\label{eq:Eres}
 \sum_{l} g_{l}^{(ij)} I_{l}^{(ij)}
 = \int_{0}^{\infty} \frac{dE}{\pi} \:
 \frac{d\delta_{ij}}{dE} \exp\left(-\frac{E}{T} \right)
 = \hat{g}_{ij} \exp\left[ -\frac{E_{ij}(T)}{T}\right] \:
\end{equation}
with a temperature dependent
effective resonance energy $E_{ij}(T)$ in each $ij$ channel
by summing the contributions of all partial waves in an effective
total phase shift
\begin{equation}
{\label{eq:phase}}
 \delta_{ij}(E) = \sum_{l} g_{l}^{(ij)} \delta_{l}^{(ij)} \: .
\end{equation}
The effective degeneracy factor $\hat{g}_{ij}= \pm g_{0}^{(ij)}$
is chosen such that it is identical to the degeneracy factor
in the s-wave. The sign is determined by that of the integral.

\begin{figure}[t]
\begin{center}
\includegraphics[width=0.7\linewidth]{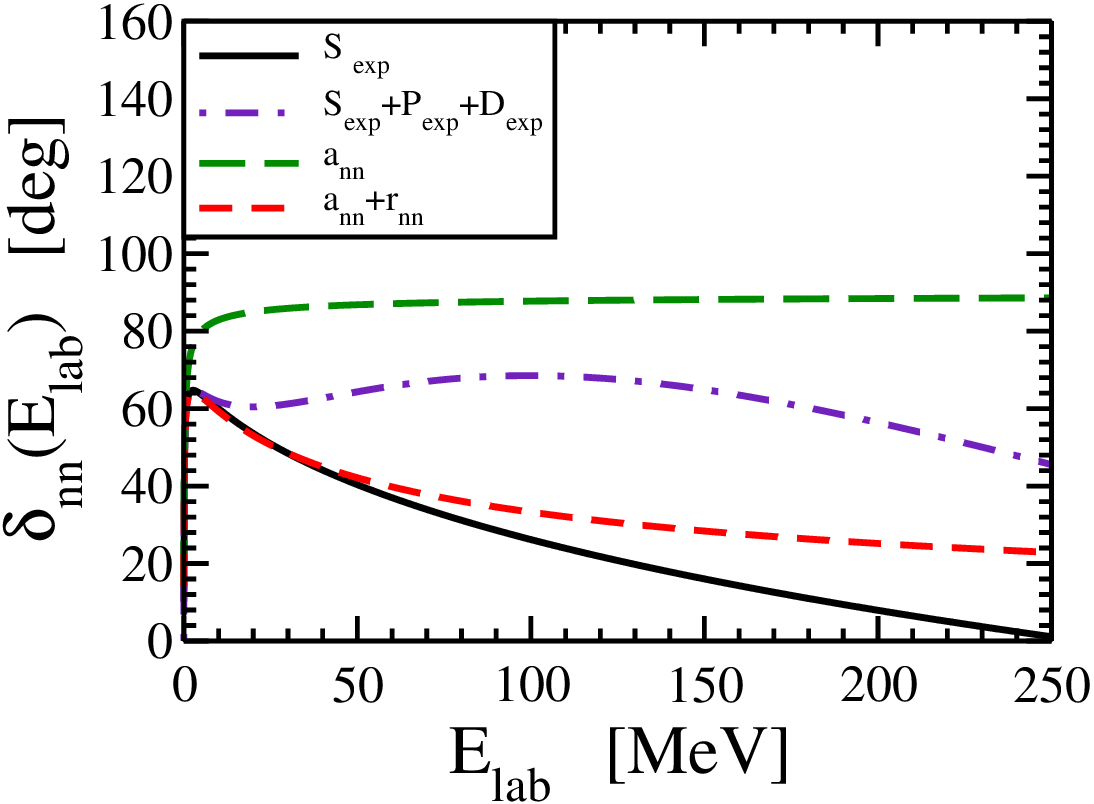}
\end{center}
\caption{\label{fig:pshifts}%
Effective phase shift $\delta_{nn}(E_{\rm lab})$ for neutron-neutron scattering
versus laboratory energy $E_{\rm lab}=2E$. Results are shown using
experimental data (including s, p and d-waves or s-waves only)
and the effective-range approximation with and without the
contribution depending on the effective range parameter, respectively.}
\end{figure}

In order to illustrate the behavior of the phase shifts and effective
resonance energies, neutron-neutron scattering is considered in the following.
In this example, the sum of the experimental phase shifts
receives contributions only from certain partial waves with
total angular momentum $J$ and
orbital angular momenta $l$ due to the Pauli principle.
Taking into account all possible partial waves with
$l=0$, $1$ and $2$, i.e. the channels
${}^{1}S_{0}$, ${}^{3}P_{0}$, ${}^{3}P_{1}$, ${}^{3}P_{2}$ and ${}^{1}D_{2}$
in spectroscopic notation ${}^{2S+1}L_{J}$, one has
\begin{equation}
\label{eq:deltatot}
 \delta_{nn}(E) 
 = \delta_{00}^{(nn)} + \delta_{01}^{(nn)} + 3 \delta_{11}^{(nn)}
 + 5 \delta_{21}^{(nn)} + 3 \delta_{12}^{(nn)}
\end{equation}
using double indices $Jl$ in $\delta_{Jl}^{(ij)}$ to indicate the
partial wave.
In Fig.~\ref{fig:pshifts} the effective phase shift for the sum of all partial
waves with $l\leq 2$ is depicted in comparison with the pure s-wave
phase shift. Since there is no experimental data on the $nn$ phase
shifts, we use the $np$ data taken from the
Nijmegen partial wave analysis \cite{Sto93} to illustrate the
comparison between 
the experimentally known phase shifts and the results using 
the effective range approximation. Note, that the laboratory
energy $E_{\rm lab}=2E$ of a neutron scattered on the target at rest
is used as the argument.
At very low energies, both curves are rapidly
rising. At energies above $\approx$~10~MeV, contributions of higher
partial waves become important.

At low temperatures, the virial integral is dominated by the
low-energy phase shifts $\delta_{l}^{(ij)}$ and only the s-wave contributes
significantly due to the large derivative with respect to the energy.
The effective-range approximation \cite{Bru96}
for the s-wave phase shift
\begin{equation}
 k \cot \delta_{0}^{(ij)} = - \frac{1}{a_{ij}} + \frac{1}{2} r_{ij} k^{2}
\end{equation}
with 
momentum $k = \sqrt{2\mu_{ij}E}$ and effective mass
$\mu_{ij} = m_{i}m_{j}/(m_{i}+m_{j})$ can be used to specify the
energy dependence of $\delta_{0}^{(ij)}$ with help of the
scattering length $a_{ij}$ and the effective range $r_{ij}$.

Fig.\ \ref{fig:pshifts} shows the result for the s-wave phase shift using only the
scattering length or both the scattering length and the effective
range parameter. The experimental phase shift is nicely reproduced
by the effective range expansion at very low energies. The
contribution of the $k^{2}$ term with $r_{ij}$ is essential to obtain
a decrease of the s-wave phase shift at higher energies. In this case, the result
follows more closely the experimental data, however,
differences with respect to the total effective phase shift
$\delta_{nn}(E)$ remain.

\begin{figure}[t]
\begin{center}
\includegraphics[width=0.7\linewidth]{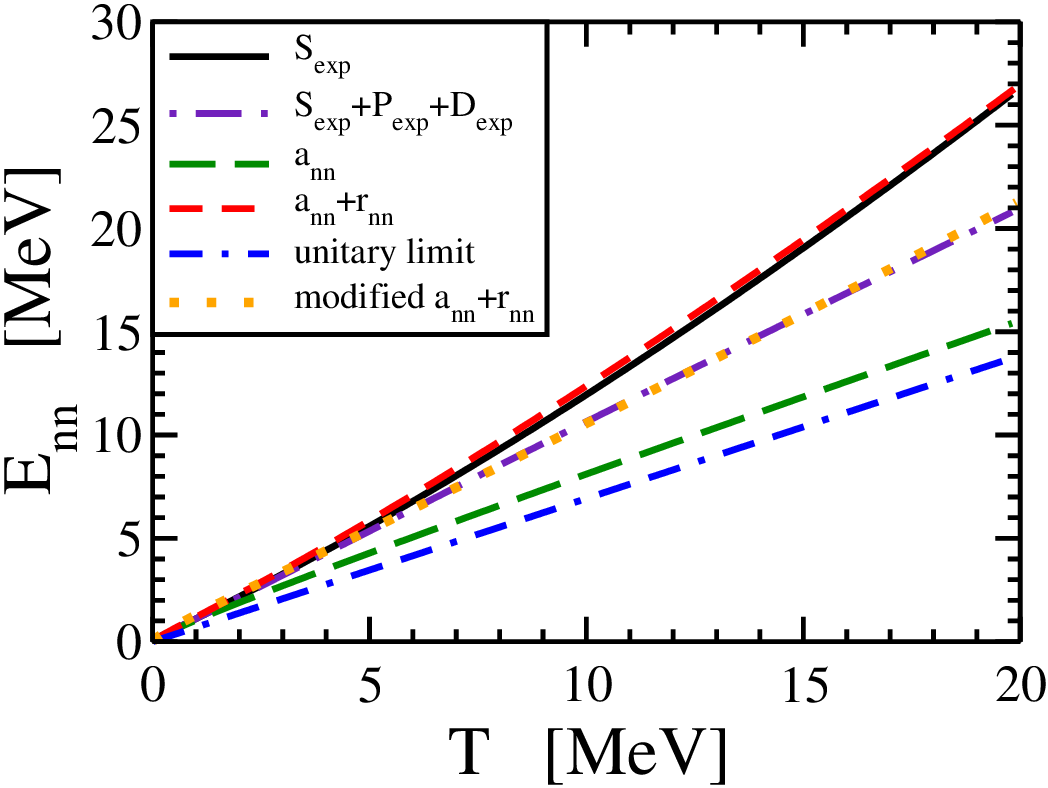}
\end{center}
\caption{\label{fig:Eres}%
Temperature dependence of the effective resonance energy $E_{nn}$ for
neutron-neutron scattering
in different approximations. See caption of Figure \ref{fig:pshifts}
and text for details.}
\end{figure}

The effective resonance energy $E_{nn}$ for $nn$
scattering derived from the phase shifts in Eq.\ (\ref{eq:Eres})
is compared for different approximations in Fig.~\ref{fig:Eres}. In general,
$E_{nn}$ rises smoothly with increasing temperature indicating that
it is not dominated by a particular resonance. Contributions
of higher partial waves reduce the effective resonance energy as
compared to the pure s-wave result.

The s-wave virial integral can be
obtained analytically in the effective-range approximation
as
\begin{eqnarray}
\label{eq:I0alrl}
 I_{0}^{(ij)}(T) & = & - \frac{1}{2} \sum_{\eta = \pm 1}
 \frac{B_{ij}^{(\eta)}}{\sqrt{A_{ij}^{(\eta)}}}
 \exp\left[\frac{1}{2\mu_{ij}TA_{ij}^{(\eta)}}\right]
 \mbox{erfc} \left[\frac{1}{\sqrt{2\mu_{ij}TA_{ij}^{(\eta)}}}\right]
\end{eqnarray}
for $2r_{ij}/a_{ij} \leq 1$.
It depends on the coefficients
\begin{equation}
 A_{ij}^{(\eta)}  =  \frac{a_{ij}^{2}}{2} \left(1-\frac{r_{ij}}{a_{ij}}
 + \eta \sqrt{1-2 \frac{r_{ij}}{a_{ij}}}  \right) 
\end{equation}
and
\begin{equation}
  B_{ij}^{(\eta)}  =   \frac{a_{ij}}{2} \left(1
 + \eta \sqrt{1-2 \frac{r_{ij}}{a_{ij}}}  \right) \: .
\end{equation}
For $r_{ij}=0$ the integral reduces to
\begin{eqnarray}
\label{eq:I0al}
 I_{0}^{(ij)}(T) & = & -\frac{a_{ij}}{2|a_{ij}|}
 \exp\left(\frac{1}{2\mu_{ij}Ta_{ij}^{2}}\right)
 \mbox{erfc} \left(\frac{1}{\sqrt{2\mu_{ij}Ta_{ij}^{2}}}\right) \: .
\end{eqnarray}
Using these results, the effective resonance energy can be
expressed explicitly as a function of the temperature and the effective-range
parameters. We use $a_{nn} = -18.818$~fm and $r_{nn}=2.834$~fm from
Ref.\ \cite{Wir01}.
As shown in Figure \ref{fig:Eres}, $E_{nn}$ in the effective range
approximation rises more
strongly with temperature when the contribution of the $r_{nn}$ term
is taken into account. At high temperatures there remains a
difference to the exact result that
takes higher partial waves into account. However, by a readjustment of
the scattering length to $a_{nn} = -11.31$~fm and the effective range
to $r_{nn}=1.06$~fm, the correct effective resonance energy
obtained with all partial waves is very well reproduced for the
relevant temperatures with the modified parameters (denoted by
modified $a_{nn}+r_{nn}$ in Figure \ref{fig:Eres}). Of course,
these modified values for $a_{nn}$ and $r_{nn}$ are not physical and do not
describe the scattering in a particular partial wave. They are only an
effective means to describe the temperature dependence of the
effective resonance energy.

There are several interesting limits for the virial integral.
For low temperatures
$T \ll A_{ij}^{(\eta)}/(2\mu_{ij})$ one finds
\begin{equation}
\label{eq:I0a}
 I_{0}^{(ij)}(T) \to - a_{ij} \sqrt{\frac{\mu_{ij}T}{2\pi}} + \dots
\end{equation}
(independent of the effective range $r_{ij}$).
In the unitary limit $a_{ij} \to - \infty$ and $r_{ij} = 0$, that
describes the situation where a bound state/resonance is just at the continuum threshold,
$ I_{0}^{(ij)}(T) = 1/2$ and the effective resonance energy is given
by $E_{ij} = T \ln 2$. This result is also depicted in Fig.~\ref{fig:Eres}.
For positive scattering length $a_{ij}$, e.g.\ in the deuteron
channel,
the integral is negative
and compensates part of the correlation strength that is located in
the bound state of the same channel.

In Ref.\ \cite{Hor06b} also the $\alpha$-particle was considered as a
fundamental constituent in the VEoS and the additional virial
coefficients $b_{\alpha}$, $b_{n\alpha}$, $b_{p\alpha}$ and $b_{\alpha\alpha}$
appeared in the formalism. (For a comparison of the virial
coefficients in the two formulations see \ref{sec:HS}.)
In our approach, the corresponding
contributions to $\Omega$ represent four-, five- and eight-body
correlations that will become irrelevant in the low-density limit as
compared
to two-nucleon correlations. Since chemical equilibrium requires
$\mu_{\alpha} = 2\mu_{n}+2\mu_{p}+B_{\alpha}$ with the
$\alpha$-particle binding energy $B_{\alpha} = 28.296$~MeV \cite{Aud03},
the $\alpha$-particle fugacity $z_{\alpha}= \exp(\mu_{\alpha}/T)$ is not
independent of the neutron and proton fugacities. The VEoS with
neutrons, protons and $\alpha$-particles was further extended in
Ref.\ \cite{OCo07} to include ${}^{3}$H and ${}^{3}$He as constituent
particles corresponding to three-body correlations. Care has to be
taken in order to avoid a double counting of states since, e.g.,
the $\alpha$-particle states can be obtained in the ${}^{3}$H-p and
${}^{3}$He-n channels (or the neglected ${}^{2}$H-${}^{2}$H channels)
that all represent four-body correlations when
nucleons are assumed to be the only fundamental particles.

\subsection{Relation to the nuclear statistical equilibrium approach}
\label{sec:NSE}

In principle, the virial expansion
could be extended to third, fourth, \dots order to include more and more
many-body correlations. The second, third, \dots virial coefficients
will contain bound state contributions that correspond to ground and
excited states of heavy nuclei. Neglecting the many-body continuum
correlations, the grand
canonical
potential becomes a sum
\begin{equation}
 \Omega = - T V \sum_{(A,Z)} \frac{g_{A,Z}}{\lambda_{A,Z}^{3}}
 \exp\left( \frac{\mu_{A,Z}}{T}\right)
\end{equation}
over all nuclei with mass number $A$, charge number $Z$, chemical
potential $\mu_{A,Z} = (A-Z) \mu_{n}+Z\mu_{p}$
and thermal wavelength $\lambda_{A,Z} =
\sqrt{2\pi/(m_{A,Z}T)}$.
The effective degeneracy factor
\begin{equation}
g_{A,Z}(T) = g_{A,Z}^{(0)} + \sum_{k} g_{A,Z}^{(k)} \exp\left( -\frac{E_{k}}{T}\right)
\end{equation}
is a sum over the ground state $0$ and all excited states $k$ of the nucleus
${}^{A}$Z with excitation energies $E_{k}$,
total angular momenta $J_{k}$ and $g_{A,Z}^{(k)} =
(2J_{k}+1)$. The summation over individual states is often replaced by an integral over
energy with an appropriate level density, see, e.g.\, Ref.\ \cite{Hem10}.
The temperature dependence of the degeneracy factor is a simple and
efficient means to include excited states of nuclei. It will be used
below in the formulation of the gRMF approach.

\subsection{Generalized Beth-Uhlenbeck approach}
\label{sec:GBU}

The description of matter in the conventional VEoS is limited to
rather low densities. However, the approach can be generalized in order to
reach higher densities and to include the effects of
dissolution of composite particles,  i.e.\ the Mott effect.
This generalized Beth-Uhlenbeck (gBU) approach
is based on a quantum statistical description with thermodynamic
Greens functions.
For details see Ref.\ \cite{Sch90}. The grand
canonical
 potential
$\Omega$ in the gBU approach assumes a similar form as in the VEoS.
The constituents are considered as quasiparticles with the correct
statistics (Fermi-Dirac or Bose-Einstein) and
with self-energies that contain already some effect of the mutual
interaction. The remaining two-body correlations (between
quasiparticles!) are contained in
a modified second virial coefficient that is given by
\begin{equation}
 b_{ij}(T) = \frac{1+\delta_{ij}}{2} \lambda_{i}^{3/2}
 \lambda_{j}^{3/2} \int dE \: f_{ij}(E+E_{\rm cont}) F_{ij}(E)
\end{equation}
with the correct distribution function $f_{ij}(E)$ of the composite system, i.e.\
Bose-Einstein for the two-nucleon-states, and the quantity\\
\begin{eqnarray}
\label{eq:Fij}
 F_{ij}(E) & = & \sum_{k} g_{k}^{(ij)} \int_{P > P_{\rm Mott}}
 \frac{d^{3}P}{(2\pi)^{3}} \: \delta(E - E_{k}^{(ij)})
 \\ \nonumber & &
 + \sum_{l} \frac{g_{l}^{(ij)}}{\pi} \int
 \frac{d^{3}P}{(2\pi)^{3}} \: 2 \sin^{2}
   \delta_{l}^{(ij)} \frac{d\delta_{l}^{(ij)}}{dE}
\end{eqnarray}
that is related to the in-medium density of states. The main
differences as compared to Eqs.\ (\ref{eq:vircoeff}) and
(\ref{eq:Dij}) are caused by medium effects. The properties of a
two-body system $ij$ depend on the total c.m.\ momentum $\vec{P}$
of the state
with respect to the medium. Hence, there is an additional summation
(integration)
over $\vec{P}$. Bound states only appear for $P$ larger than the Mott
momentum $P_{\rm Mott}$ because at lower total momenta their formation is suppressed
by the action of the Pauli principle. The phase shifts
$\delta_{l}^{(ij)}(P,T,\mu_{i},\mu_{j})$ are determined by the
in-medium T-matrix describing the scattering of two quasiparticles. Similarly
$E_{k}^{(ij)}(P,T,\mu_{i},\mu_{j})$ is the medium-dependent energy of a bound state.
Energies in Eq.\ (\ref{eq:Fij}) are measured with respect to the
continuum edge $E_{\rm cont}$ that corresponds to the energy in the
frame of the medium where the relative momentum of the scattering particles is zero.
The $2 \sin^{2} \delta_{l}^{(ij)}$ factor takes into account that part of
interaction effects is already contained in the self-energies of the
quasiparticles. The second term in Eq.\ (\ref{eq:Fij}) is related to the
number of correlated quasiparticles that is different to the number
of correlated particles as calculated by the usual Beth-Uhlenbeck
approach in Eq.\ (\ref{eq:Dij}).

\section{Generalized relativistic mean-field model}
\label{sec:gRMF}
\subsection{General formalism}

In Ref.\ \cite{Typ10} a RMF model with
density-dependent meson-nucleon couplings was generalized to include
light clusters (bound states of ${}^{2}$H, ${}^{3}$H, ${}^{3}$He,
${}^{4}$He) as additional degrees of freedom with medium-dependent
binding energies. These clusters couple minimally to
the meson fields with a strength that was assumed to be a multiple of the nucleon
coupling strength. Scattering states were not considered in the RMF
model of in Ref.\ \cite{Typ10}.
The approach can easily be extended when more than the above mentioned nuclei
need to be included.
Such a model with nuclei as explicit 
degrees of freedom we call the gRMF model in this paper.

The grand canonical
potential of the model can be formulated as
a function $\Omega(T,V,\tilde{\mu}_{i},\omega,\rho,\sigma,\delta)$
depending on the temperature $T$, volume $V$, relativistic chemical
potentials $\tilde{\mu}_{i}=m_{i}+\mu_{i}$ of the particles and the
spatially constant meson fields
$\omega$, $\rho$, $\sigma$ and $\delta$
under consideration. Here, all four mesons are included  to cover
the four possible combinations of scalars and vectors in Lorentz
and isospin space. We
neglect antiparticles in the following because they give sizable
contributions only at temperatures much higher than those relevant in the
considered astrophysical applications. In the general case of
inhomogeneous matter, not discussed here, $\Omega$ becomes a functional of the spatially
varying meson fields and their gradients.


For uniform nuclear matter, the grand canonical
potential can be written as
\begin{eqnarray}
\label{eq:Omega_rmf}
 \Omega 
 & = &  \sum_{i} \Omega_{i}
 - V \left[ \frac{1}{2} \left( m_{\omega}^{2} \omega^{2}
 + m_{\rho}^{2} \rho^{2} - m_{\sigma}^{2} \sigma^{2}
 - m_{\delta}^{2} \delta^{2}
 \right)
 \right. \\ \nonumber & & \left.
 + \left( \Gamma_{\omega}^{\prime} \omega n_{\omega}
 + \Gamma_{\rho}^{\prime} \rho n_{\rho}
 - \Gamma_{\sigma}^{\prime} \sigma n_{\sigma}
 - \Gamma_{\delta}^{\prime} \delta n_{\delta} \right) \left( n_{n}+n_{p}\right)
 \right]
\end{eqnarray}
where $\Gamma_{i}'$ are the derivatives of the density dependent
couplings which determine the nucleon-meson coupling strength. 
Contributions from individual particles (nucleons and clusters) are
\begin{equation}
\label{eq:Omegai}
 \Omega_{i} = \mp g_{i} V T \int \frac{d^{3}k}{(2\pi)^{3}} \: \ln \left[ 1
 \pm \exp \left( - \frac{E_{i}-\tilde{\mu}_{i}}{T}\right)\right]
\end{equation}
where the upper (lower) signs refer to fermions (bosons). In the case
of Maxwell-Boltzmann statistics, Eq.\ (\ref{eq:Omegai}) reduces to
\begin{equation}
\label{eq:OmegaMB}
 \Omega_{i} = g_{i} V T \int \frac{d^{3}k}{(2\pi)^{3}} \:
\exp \left( - \frac{E_{i}-\tilde{\mu}_{i}}{T}\right) \: .
\end{equation}
We assume
that the degeneracy factors $g_{i}$ can depend in general on the temperature as
in the case of NSE models, cf.\ Section \ref{sec:NSE}.
The densities $n_{\omega}$, $n_{\rho}$, $n_{\sigma}$ and $n_{\delta}$
appearing in (\ref{eq:Omega_rmf}) are themselves functions of the
temperature, chemical potentials and meson fields (see below).

The mass of a cluster $i=d,t,h,\alpha,\dots$ with $N_{i}$ neutrons and
$Z_{i}$ protons is written as
\begin{equation}
\label{eq:m_i}
 m_{i} = N_{i} m_{n} + Z_{i} m_{p} -  (1-\delta_{in})(1-\delta_{ip})
 B_{i}^{\rm (vac)}
\end{equation}
with neutron and proton rest masses $m_{n}$ and $m_{p}$, respectively,
and the vacuum binding energy $B_{i}^{\rm (vac)}>0$ for composite particles.
For neutrons and protons the binding energy does not appear.
 Chemical equilibrium leads to the constraint
\begin{equation}
 \tilde{\mu}_{i} = N_{i} \tilde{\mu}_{n} + Z_{i} \tilde{\mu}_{p}
\end{equation}
for the relativistic chemical potentials of the clusters.
With every particle $i=n,p,d,t,h,\alpha,\dots$ of momentum
$\vec{k}$, an energy
\begin{equation}
 E_{i}(\vec{k}) = V_{i} + \sqrt{k^{2} + (m_{i}-S_{i})^{2}}
\end{equation}
is associated that contains the vector and scalar self-energies
$V_{i}$ and $S_{i}$, respectively. These potentials are given by
\begin{eqnarray}
 V_{i} & = &\Gamma_{i\omega} \omega +
 \Gamma_{i\rho}  \rho + (\delta_{in} + \delta_{ip})
 V^{(r)} \: ,
 \\
\label{eq:Si}
 S_{i} & = & \Gamma_{i\sigma}  \sigma +
 \Gamma_{i\delta} \delta
 - (1-\delta_{in})(1-\delta_{ip})\Delta B_{i}
\end{eqnarray}
with meson fields $m=\omega,\rho,\sigma,\delta$ and
their couplings $\Gamma_{im}$ to the particle $i$.
We assume that the coupling strength of the clusters is a multiple
of that of the nucleons, i.e.\
\begin{equation}
\label{eq:Gim}
 \Gamma_{im} = g_{im} \Gamma_{m}(\varrho)
\end{equation}
with
\begin{eqnarray}
\label{eq:gim}
 & & g_{i\omega}  = g_{i\sigma} = N_{i}+Z_{i} \: ,
 \\
 & & g_{i\rho} = g_{i\delta} = N_{i} - Z_{i} \: .
\end{eqnarray}
Other choices are possible and should be explored in the future.
The meson-nucleon couplings $\Gamma_{m}(\varrho)$ depend on the total nucleon
density $\varrho = n_{n}+n_{p}$ only. The functional form and the 
parameters of the density dependent couplings $\Gamma_{m}(\varrho)$
used in the gRMF model were chosen in such a way to describe the 
properties of finite nuclei and nuclear matter parameters at saturation density.
The vector self-energy contains the rearrangement term
\begin{equation}
\label{eq:vr}
 V^{(r)} = \Gamma_{\omega}^{\prime}  n_{\omega}\omega
 + \Gamma_{\rho}^{\prime}n_{\rho} \rho
 - \Gamma_{\sigma}^{\prime}  n_{\sigma}\sigma
 - \Gamma_{\delta}^{\prime} n_{\delta}\delta
\end{equation}
that is only relevant for nucleons.
In the medium, the actual binding energy of a cluster $B_{i}=
B_{i}^{\rm (vac)}-\Delta B_{i}$ will be shifted
with respect to the vacuum value $B_{i}^{\rm (vac)}$ by a quantity
$\Delta B_{i}$.
This medium-dependent binding
energy shift $\Delta B_{i}$ for composite particles
is parametrized as a function
of temperature and the Lorentz vector meson fields $\omega$ and $\sigma$,
see, e.g.\ Ref.\ \cite{Typ10}.
The specific form of these shifts is not important in the present
discussion of the low-density behavior of the EoS but it affects the
transition to high densities, see Section \ref{sec:high}.

The source densities in Eqs.\ (\ref{eq:Omega_rmf}) and (\ref{eq:vr})
\begin{eqnarray}
 n_{\omega} & = & \sum_{i} g_{i\omega} n_{i} \: ,
 \quad
 n_{\rho}  =  \sum_{i} g_{i\rho} n_{i} \: ,
 \\
 n_{\sigma} & = & \sum_{i} g_{i\sigma} n_{i}^{(s)} \: ,
 \quad
 n_{\delta} =  \sum_{i} g_{i\delta} n_{i}^{(s)}
\end{eqnarray}
depend on the vector and scalar particle number densities $n_{i}$ and
$n_{i}^{(s)}$, respectively, that are defined as
\begin{eqnarray}
\label{eq:ni_rmf}
 n_{i} & = & - \left. \frac{\partial \Omega}{\partial \tilde{\mu}_{i}}
 \right|_{T,V,\tilde{\mu}_{j\neq i}}
 = g_{i} \int \frac{d^{3}k}{(2\pi)^{3}} \: f_{i} \: ,
 \\
 n_{i}^{(s)} & = & g_{i} \int \frac{d^{3}k}{(2\pi)^{3}} \: f_{i}
 \frac{m_{i}-S_{i}}{\sqrt{k^{2}-(m_{i}-S_{i})^{2}}}
\end{eqnarray}
with the Fermi-Dirac (Bose-Einstein) distribution function
\begin{equation}
\label{eq:FD}
 f_{i}(\vec{k}) = \left[\exp\left( \frac{E_{i}-\tilde{\mu}_{i}}{T}\right) \pm 1
\right]^{-1}
\end{equation}
or the Maxwell-Boltzmann distribution function
\begin{equation}
\label{eq:MB}
 f_{i}(\vec{k}) = \exp\left( -
     \frac{E_{i}-\tilde{\mu}_{i}}{T} \right)
\end{equation}
depending on the particle statistics.

The field equations for the mesons are found from the functional $\Omega$
with the help of
the Euler-Lagrange equations. They have the form
\begin{eqnarray}
\label{eq:feq_o}
 m_{\omega}^{2} \omega & = & \Gamma_{\omega} n_{\omega}
 + \sum_{i} (1-\delta_{in})(1-\delta_{ip})
 n_{i}^{(s)} \frac{\partial \Delta B_{i}}{\partial\omega}   \: ,
 \\
\label{eq:feq_r}
 m_{\rho}^{2} \rho & = & \Gamma_{\rho} n_{\rho}
 + \sum_{i} (1-\delta_{in})(1-\delta_{ip})
 n_{i}^{(s)}  \frac{\partial \Delta B_{i}}{\partial\rho}  \: ,
 \\
\label{eq:feq_s}
 m_{\sigma}^{2} \sigma & = & \Gamma_{\sigma} n_{\sigma} \: ,
 \\
\label{eq:feq_d}
 m_{\delta}^{2} \delta & = & \Gamma_{\delta} n_{\delta}
\end{eqnarray}
with additional contributions to the source terms
due to the dependence of the binding energies $B_{i}$ on the vector meson fields.
Finally, the entropy is obtained as
\begin{eqnarray}
\label{eq:entropy}
 S & = & - V \sum_{i} g_{i} \int \frac{d^{3}k}{(2\pi)^{3}} \:
 \left[ f_{i} \ln f_{i} \pm \left( 1 \mp f_{i} \right)
 \ln \left( 1 \mp f_{i} \right) \right]
 \\ & & \nonumber
 - V \sum_{i} \left( 1- \delta_{in} \right) \left( 1- \delta_{ip}
   \right)  n_{i}^{(s)} \frac{\partial \Delta B_{i} }{\partial T}
 - \sum_{i} \frac{\Omega_{i}}{g_{i}} \frac{dg_{i}}{dT}
\end{eqnarray}
with the usual single-particle contribution, a term due to the
temperature dependence of the cluster binding energy shifts
$\Delta B_{i}$ and a contribution caused by
internal excitations of the clusters, i.e.\
the introduction of the temperature dependence of the degeneracy factors,
described in Sect.\ \ref{sec:ExtRMF}.
In case of the Maxwell-Boltzmann
statistics for a particle $i$, the integrand contains only the term
$f_{i} \ln f_{i}$. 

\subsection{Scheme of the fugacity expansion}

Introducing the modified fugacity
\begin{equation}
 \tilde{z}_{i} = \exp \left( \frac{ \tilde{\mu}_{i} -m_{i}
 +S_{i}-V_{i}}{T}\right)
 = \exp \left( \frac{S_{i}-V_{i}}{T} \right) z_{i}
\end{equation}
we write the distribution function as
\begin{equation}
 f_{i} = \left[\tilde{z}_{i}^{-1} \exp\left(
 \frac{e_{i}}{T}\right) \pm 1
\right]^{-1} = \frac{\tilde{z}_{i} \exp\left( -
 \frac{e_{i}}{T}\right)}{1 \pm \tilde{z}_{i} \exp\left( -
 \frac{e_{i}}{T}\right)}
\end{equation}
with the kinetic energy $e_{i}(k) = \sqrt{k^{2}+(m_{i}-S_{i})^{2}}-(m_{i}-S_{i})\ge 0$.
For $\tilde{z}_{i}\ll 1$ the distribution function
can be expanded in a power series in
$\tilde{z}_{i}\exp\left(-e_{i}/T\right)$.
The appearing momentum space integrals can be evaluated explicitly \cite{Joh96}
with the integral representation of the modified Bessel functions $K_{\nu}(x)$
\cite{Abr65}. For
the contributions (\ref{eq:Omegai}) of the individual particles to the grand
canonical
potential we find
\begin{eqnarray}
\label{eq:Omegai_2}
 \lefteqn{\Omega_{i} =
 - VT \frac{g_{i}}{\lambda_{i}^{3}}
 \left(\frac{m_{i}-S_{i}}{m_{i}}\right)^{3/2}}
 \\ \nonumber & & \times
\sum_{n=0}^{\infty} \frac{(\mp 1)^{n}}{(n+1)^{5/2}}
 k_{2} \left[ (n+1) \frac{m_{i}-S_{i}}{T} \right]
 \exp \left[ (n+1) \frac{S_{i}-V_{i}}{T} \right] z_{i}^{n+1}
\end{eqnarray}
with the non-relativistic fugacity $z_{i}$
and the functions
\begin{equation}
 k_{\nu}(x) = \sqrt{\frac{2x}{\pi}} \exp(x) K_{\nu}(x) \: .
\end{equation}
Similarly, the vector density is obtained as
\begin{eqnarray}
\label{eq:ni_2}
 \lefteqn{n_{i} =
   \frac{g_{i}}{\lambda_{i}^{3}}
 \left(\frac{m_{i}-S_{i}}{m_{i}}\right)^{3/2}}
 \\ \nonumber & & \times
\sum_{n=0}^{\infty} \frac{(\mp 1)^{n}}{(n+1)^{3/2}}
 k_{2} \left[ (n+1) \frac{m_{i}-S_{i}}{T} \right]
 \exp \left[ (n+1) \frac{S_{i}-V_{i}}{T} \right] z_{i}^{n+1} \: .
\end{eqnarray}
For the scalar density $n_{i}^{(s)}$, $k_{2}$ has to be replaced with $k_{1}$.
In case of Maxwell-Boltzmann statistics, only the $n=0$ term remains.
Without interaction, i.e.\ $S_{i} = V_{i}=0$, the results for a
relativistic Fermi or Bose gas are recovered. For $T \to 0$,
i.e.\ $x = (n+1)(m_{i}-S_{i})/T\to \infty$, one finds from
the asymptotic expansion \cite{Abr65}
\begin{equation}
\label{eq:k_asym}
 k_{\nu}(x) = 1 + \frac{\mu-1}{8x} + \frac{(\mu-1)(\mu-9)}{2!(8x)^{2}}
+ \frac{(\mu-1)(\mu-9)(\mu-25)}{3!(8x)^{3}} + \dots
\end{equation}
with $\mu = 4\nu^{2}$ the common expressions
\begin{equation}
 \Omega_{i} = - VT \frac{g_{i}}{\lambda_{i}^{3}}
 \sum_{m=0}^{\infty} \frac{(\mp 1)^{m}}{(j+1)^{5/2}} z_{m}^{m+1} \: ,
\end{equation}
and
\begin{equation}
 n_{i} = \frac{g_{i}}{\lambda_{i}^{3}}
 \sum_{m=0}^{\infty} \frac{(\mp 1)^{m}}{(m+1)^{3/2}} z_{m}^{m+1}
\end{equation}
for non-relativistic particles.
In this limit, $n_{i}^{(s)} = n_{i}$, because the scalar density differs 
from the baryon density only in the relativistic description.
Since the scalar and vector self-energies are themselves functions of
the densities, a second expansion of the 
series (\ref{eq:Omegai_2}) and (\ref{eq:ni_2}) is required.

\subsection{Fugacity expansion of the grand canonical
potential up to second order}

In order to compare the
true series expansion of the grand canonical potential $\Omega$
of the gRMF model with the form
(\ref{eq:Omega_veos}) in the VEoS approach at low densities, we expand
(\ref{eq:Omega_rmf}) up to second order in the fugacities of neutrons
and protons. In the following we will only consider neutrons, protons
and deuterons in the density expansion.
Nuclei with mass numbers $A\geq 3$ do not contribute in the second 
order of expansion since we only consider
nucleons as basic constituents.
The explicit contribution
of the deuteron ground state appears in the two-nucleon correlation term
with
\begin{equation}
 z_{d} = z_{n} z_{p} \exp\left(\frac{B_{d}}{T}\right) \: .
\end{equation}

For sufficiently low nucleon densities, the self-energies 
of the nucleons are approximately linear in the
nucleon densities.
Hence, the contribution of the individual nucleon can be approximated as
\begin{eqnarray}
\label{eq:Omega_i}
   \Omega_{i} & \approx & - VT \frac{g_{i}}{\lambda_{i}^{3}}
 \left\{ k_{2} \left( \frac{m_{i}}{T} \right) \left[ 1 +
     \frac{S_{i}}{T} \left( 1-
 \frac{k_{2}^{\prime} \left( \frac{m_{i}}{T} \right)}{k_{2}
 \left( \frac{m_{i}}{T} \right)}
 - \frac{3}{2} \frac{T}{m_{i}} \right) - \frac{V_{i}}{T}\right] z_{i}
 \right. \\ \nonumber & & \left.
 - \frac{1}{2^{5/2}} k_{2} \left( 2 \frac{m_{i}}{T} \right) z_{i}^{2}
\right\} \: .
\end{eqnarray}
Applying the recursion relation
\begin{equation}
 k_{\nu}^{\prime}(x) = \left( 1 - \frac{2\nu-1}{2x}\right) k_{\nu} -
 k_{\nu-1}(x) \: ,
\end{equation}
the expression (\ref{eq:Omega_i}) for the nucleons reduces to
\begin{eqnarray}
   \Omega_{i} & \approx & - VT
 \left[ \left( 1 - \frac{V_{i}}{T}\right) x_{i}
 + \frac{S_{i}}{T} y_{i}
 - \frac{g_{i}}{2^{5/2}\lambda_{i}^{3}} k_{2}
 \left( 2 \frac{m_{i}}{T} \right) z_{i}^{2} \right]
\end{eqnarray}
with the abbreviations
\begin{eqnarray}
 x_{i} & = & \frac{g_{i}}{\lambda_{i}^{3}} k_{2} \left(
   \frac{m_{i}}{T} \right) z_{i} \: ,
 \\
  y_{i} & = & \frac{g_{i}}{\lambda_{i}^{3}} k_{1} \left(
   \frac{m_{i}}{T} \right) z_{i} \: .
\end{eqnarray}
For the deuteron contribution we have
\begin{eqnarray}
  \Omega_{d} & = & - VT \frac{g_{d}}{\lambda_{d}^{3}}
  k_{2} \left( \frac{m_{d}}{T} \right) z_{n}z_{p} \exp\left(\frac{B_{d}}{T}\right)
\end{eqnarray}
without self-energy terms which contribute only in higher order 
of the fugacities. Contrary the nucleon self-energies contribute already in the
lowest order of the fugacities
\begin{eqnarray}
 V_{n} & \approx &
 C_{\omega} (x_{n} + x_{p}) + C_{\rho} (x_{n} - x_{p}),
 \\
 V_{p} & \approx &
 C_{\omega} (x_{n} + x_{p}) - C_{\rho} (x_{n} - x_{p}),
 \\
 S_{n} & \approx &
 C_{\sigma} (y_{n}+y_{p}) + C_{\delta} (y_{n} - y_{p}),
 \\
 S_{p} & \approx &
 C_{\sigma} (y_{n}+y_{p}) - C_{\delta} (y_{n} - y_{p})
\end{eqnarray}
with coefficients
\begin{equation}
\label{eq:Cm}
 C_{m} = \frac{\Gamma_{m}^{2}(0)}{m_{m}^{2}}
\end{equation}
for $m=\omega,\rho,\sigma,\delta$ that depend on the density dependent meson couplings
$\Gamma_{m}(\varrho)$ taken in the limit $\varrho\rightarrow0$.
The rearrangement term in the vector self-energy does not contribute
at this level because it is at least quadratic in the densities. With
the help of the field equations (\ref{eq:feq_o})-(\ref{eq:feq_d}), 
the mesonic contributions in
(\ref{eq:Omega_rmf}) can be expressed
as quadratic forms of the nucleon densities. Finally, after performing all necessary
 expansions up to second order in the nucleon fugacities, we obtain the resulting
 grand canonical potential
 \begin{eqnarray}
\label{eq:Omega_rmf2}
 \lefteqn{\Omega(T,V,\mu_{n},\mu_{p})=}
 \\ \nonumber & &
 - TV \frac{g_{n}}{\lambda_{n}^{3}}  \left[ k_{2} \left( \frac{m_{n}}{T}
 \right) z_{n}
 - \frac{1}{2^{5/2}} k_{2}\left( 2\frac{m_{n}}{T}\right)
 z_{n}^{2} \right]
 \\ \nonumber & &
  - TV \frac{g_{p}}{\lambda_{p}^{3}}  \left[ k_{2} \left( \frac{m_{p}}{T}
 \right) z_{p}
 - \frac{1}{2^{5/2}} k_{2}^{2}\left( 2\frac{m_{p}}{T}\right)
 z_{p}^{2} \right]
 \\ \nonumber & &
  - TV \frac{g_{d}}{\lambda_{d}^{3}} k_{2}\left(\frac{m_{d}}{T}\right)
 \exp \left( \frac{B_{d}}{T}\right)
 z_{n}z_{p}
 \\ \nonumber & &
 + \frac{V}{2} \frac{g_{n}^{2}}{\lambda_{n}^{6}}
 \left[ \left( C_{\omega}  + C_{\rho} \right)
   k_{2}^{2}\left( \frac{m_{n}}{T}\right)
 -  \left( C_{\sigma}  + C_{\delta} \right)
  k_{1}^{2}\left(
   \frac{m_{n}}{T}\right) \right]
  z_{n}^{2}
 \\ \nonumber & &
 + \frac{V}{2} \frac{g_{p}^{2}}{\lambda_{p}^{6}}
 \left[ \left( C_{\omega}  + C_{\rho} \right)
   k_{2}^{2}\left( \frac{m_{p}}{T}\right)
 -  \left( C_{\sigma}  + C_{\delta} \right)
  k_{1}^{2}\left(
   \frac{m_{p}}{T}\right) \right]
 z_{p}^{2}
 \\ \nonumber & &
 + V \frac{g_{n}g_{p}}{\lambda_{n}^{3}\lambda_{p}^{3}}
 \left[ \left( C_{\omega}  - C_{\rho} \right)
 k_{2}\left( \frac{m_{n}}{T}\right) k_{2}\left( \frac{m_{p}}{T}\right)
 \right. \\ \nonumber & & \left.
 - \left( C_{\sigma}  - C_{\delta} \right)
 k_{1}\left( \frac{m_{n}}{T}\right) k_{1}\left( \frac{m_{p}}{T}\right)
 \right]
 z_{n}z_{p}
\end{eqnarray}
with contributions from free nucleons and their correlation
due to statistics, the deuteron bound state and
the interaction.

\subsection{Comparison of fugacity expansions}
\label{sec:comp}
A comparison of Eq. (\ref{eq:Omega_veos}) with the corresponding
expansion (\ref{eq:Omega_rmf2}) permits to extract the virial
coefficients $b_{n}$, $b_{p}$, $b_{nn}$, $b_{pp}$ and $b_{np}$ of the
gRMF model. The first-order coefficients receive a
relativistic correction with
\begin{eqnarray}
 b_{n} & = & g_{n} k_{2} \left( \frac{m_{n}}{T} \right) \: ,
\\
 b_{p} & = & g_{p} k_{2} \left( \frac{m_{p}}{T} \right) \: .
\end{eqnarray}
They depend on the temperature now. In the non-relativistic limit 
$T/m_{i}\to0$ the correction will vanish due to the
asymptotic expansion of the function $k_{2}$, cf.\ Eq. (\ref{eq:k_asym}).
Below we will show that these corrections are important and should be taken into
account for an improved description of low-density nuclear matter.
Similarly, there is a relativistic
modification to the statistical corrections in $b_{nn}$ and $b_{pp}$,
i.e.\ Eqs. (\ref{eq:bnn}) and (\ref{eq:bpp}), respectively, and
to the deuteron contribution to (\ref{eq:bnp0}).
From the comparison of (\ref{eq:Omega_veos}) and (\ref{eq:Omega_rmf2}), three independent
relations for the channels $nn$, $pp$ and $np$ remain
\begin{eqnarray}
\label{eq:nn_comp}
 \lefteqn{-\frac{T}{\lambda_{n}^{3}} \left[b_{nn}
 + \frac{g_{n}}{2^{5/2}} k_{2} \left( 2\frac{m_{n}}{T}\right) \right]
 = -\frac{T}{\lambda_{nn}^{3}}
 k_{2}\left( \frac{2m_{n}}{T}\right)
 \sum_{l} g_{l}^{(nn)} I_{l}^{(nn)}}
 \\ \nonumber & = &
 \frac{1}{2} \frac{g_{n}^{2}}{\lambda_{n}^{6}}
 \left[ \left( C_{\omega}  + C_{\rho} \right)
   k_{2}^{2}\left( \frac{m_{n}}{T}\right)
 -  \left( C_{\sigma}  + C_{\delta} \right)
  k_{1}^{2}\left(
   \frac{m_{n}}{T}\right) \right]  \: ,
 \\
\label{eq:pp_comp}
\lefteqn{-\frac{T}{\lambda_{p}^{3}} \left[ b_{pp}
 + \frac{g_{p}}{2^{5/2}} k_{2} \left( 2\frac{m_{p}}{T}\right) \right]
 = -\frac{T}{\lambda_{pp}^{3}}
 k_{2}\left( \frac{2m_{p}}{T}\right)
 \sum_{l} g_{l}^{(pp)} I_{l}^{(pp)}}
 \\ \nonumber & = &
 \frac{1}{2} \frac{g_{p}^{2}}{\lambda_{p}^{6}}
 \left[ \left( C_{\omega}  + C_{\rho} \right)
   k_{2}^{2}\left( \frac{m_{p}}{T}\right)
 -  \left( C_{\sigma}  + C_{\delta} \right)
  k_{1}^{2}\left(
   \frac{m_{p}}{T}\right) \right]  \: ,
 \\
\label{eq:np_comp}
 \lefteqn{-T\left[ 2\frac{b_{np}}{\lambda_{n}^{3/2}\lambda_{p}^{3/2}}
 - \frac{g_{d}}{\lambda_{d}^{3}} k_{2}\left( \frac{m_{d}}{T}\right)
 \exp \left( \frac{B_{d}}{T}\right)\right]}
 \\ \nonumber & = &
 -\frac{T}{\lambda_{np}^{3}}
 k_{2}\left( \frac{m_{n}+m_{p}}{T}\right)
 \sum_{l} \left[ g_{l}^{(np1)} I_{l}^{(np1)}
 + g_{l}^{(np0)} I_{l}^{(np0)} \right]
 \\ \nonumber & = &
 \frac{g_{n}g_{p}}{\lambda_{n}^{3}\lambda_{p}^{3}}
 \left[ \left( C_{\omega}  - C_{\rho} \right)
 k_{2}\left( \frac{m_{n}}{T}\right) k_{2}\left( \frac{m_{p}}{T}\right)
 \right. \\ \nonumber & & \left.
 - \left( C_{\sigma}  - C_{\delta} \right)
 k_{1}\left( \frac{m_{n}}{T}\right) k_{1}\left( \frac{m_{p}}{T}\right)
 \right]
\end{eqnarray}
that connect the virial integrals with the strengths $C_m$ of the zero-density
meson-nucleon couplings  $\Gamma_{m}(0)$ of the gRMF model.
The expected correction factors $k_{2}(m_{i}/T)$ 
due to the relativistic effects were added to the VEoS part 
(expressed through virial integrals $I_{l}^{ij}$)
of relations (\ref{eq:nn_comp}) to (\ref{eq:np_comp}).

\subsection{Temperature independent limit of consistency conditions}
\label{sec:T0limit}

The consistency relations (\ref{eq:nn_comp}) to (\ref{eq:np_comp})
were derived from a comparison of two fugacity expansions that
are valid at low densities and finite temperatures, i.e. in the virial
limit.
These conditions arise from the contributions in the fugacity
expansion that are quadratic in the fugacities.
We perform an expansion of Eqs.\ (\ref{eq:nn_comp}) to
(\ref{eq:np_comp}) in powers of small $T/m_{i}$ and keep the lowest-order terms which are independent of $T$. 
This is equivalent of taking the limit  $T\to 0$ in
(\ref{eq:nn_comp}) to (\ref{eq:np_comp}).
In this limit there are no
relativistic corrections, only s-wave
contributions are relevant and the limit (\ref{eq:I0a}) of the integral with the
scattering lengths can be used.
In the gRMF model the nucleon-nucleon interaction is
identical in the $nn$ and $pp$ systems. Thus, the first two equations,
(\ref{eq:nn_comp}) and (\ref{eq:pp_comp}) cannot be considered
independent but should be combined.
Denoting the scattering lengths in
the s-wave channels explicitly with $a^{(nn)}_{{}^{1}S_{0}}$,
$a^{(pp)}_{{}^{1}S_{0}}$, $a^{(np)}_{{}^{1}S_{0}}$ and
$a^{(np)}_{{}^{3}S_{1}}$
and considering the degeneracy factors $g_{n} = g_{p} = 2$,
$g^{(nn)}_{{}^{1}S_{0}}=g^{(pp)}_{{}^{1}S_{0}}=g^{(np)}_{{}^{1}S_{0}}=1$ and
$g^{(np)}_{{}^{3}S_{1}}=3$,
the two consistency conditions
\begin{equation}
\label{eq:con1}
  C_{\omega} - C_{\sigma} =
   \pi \left\{ \frac{1}{2} \left[ \frac{a^{(nn)}_{{}^{1}S_{0}}}{m_{n}}
 + \frac{a^{(pp)}_{{}^{1}S_{0}}}{m_{p}} \right]
  + \frac{m_{n}+m_{p}}{m_{n}m_{p}}
  \frac{a^{(np)}_{{}^{1}S_{0}} + 3  a^{(np)}_{{}^{3}S_{1}}}{4} \right\}
\end{equation}
and
\begin{equation}
\label{eq:con2}
  C_{\rho} - C_{\delta} =
   \pi \left\{ \frac{1}{2} \left[ \frac{a^{(nn)}_{{}^{1}S_{0}}}{m_{n}}
 + \frac{a^{(pp)}_{{}^{1}S_{0}}}{m_{p}} \right]
  - \frac{m_{n}+m_{p}}{m_{n}m_{p}}
  \frac{a^{(np)}_{{}^{1}S_{0}} + 3  a^{(np)}_{{}^{3}S_{1}}}{4} \right\}
\end{equation}
for the isovector and isoscalar meson couplings are obtained.
The left side of Eqs. (\ref{eq:con1}) to (\ref{eq:con2}) 
corresponds to the gRMF expansion while
the right side represents the virial expansion in the effective-range approximation.

\begin{table}[t]
\caption{\label{tab:rmf}%
Coupling coefficients (\ref{eq:Cm}) in the
DD2 parametrization of the  gRMF model \cite{Typ10}
and the DD-ME$\delta$ parametrization of Ref.\ \cite{Roc11}.}
\begin{center}
  \begin{tabular}{lcc}
  \hline
  Model & DD2 & DD-ME$\delta$ \\
    \hline
    meson $m$ & $C_{m}$ [fm${}^{2}$] & $C_{m}$ [fm${}^{2}$]\\
    \hline
    $\omega$  & 17.250448 & 14.539639\\
    $\sigma$  & 22.639364 & 19.443381\\
    $\rho$    & 2.483932 & 6.017331\\
    $\delta$  & $-$      & 3.462716\\
    \hline
  \end{tabular}
\end{center}
\end{table}

With the coupling constants at zero density $\Gamma_{m}(0)$ and the
meson masses $m_{m}$
the coupling coefficients $C_{m}$ can be calculated for different
RMF models. In Table \ref{tab:rmf} the results are shown for the gRMF model with
parametrization DD2 without $\delta$ meson following
Ref.\ \cite{Typ10} and for a new density dependent
RMF parametrization
DD-ME$\delta$ from \cite{Roc11} based on ab-initio calculations in
nuclear matter including a Lorentz scalar isovector $\delta$ meson.\\
In the DD2 parametrization  the differences of the coupling
coefficients are
$C_{\omega} - C_{\sigma}  =  -5.39~\mbox{fm}^{2}$
and
$C_{\rho} - C_{\delta} = 2.48~\mbox{fm}^{2}$, respectively.
In contrast, the calculation with the scattering lengths from
Ref.\ \cite{Wir01} gives
$C_{\omega} - C_{\sigma}  =  -14.15~\mbox{fm}^{2}$
and
$C_{\rho} - C_{\delta} = -9.61~\mbox{fm}^{2}$.
It is obvious that the coupling coefficients of the gRMF
model do not obey the consistency relations to guarantee the agreement
with the VEoS at low densities.

For the recently developed parametrization DD-ME$\delta$
\cite{Roc11},
the differences of the coupling coefficients
turn out to be $C_{\omega} - C_{\sigma}  =  -4.90~\mbox{fm}^{2}$
and $C_{\rho} - C_{\delta} = 2.55~\mbox{fm}^{2}$.
These numbers are close to those of the DD2 parametrization.
There are still differences to the values required by the
consistency conditions (\ref{eq:con1}) and (\ref{eq:con2}) and a more
drastic modification of the low-density couplings seems to be necessary.

In the DD2 parameter fit of the gRMF approach
a particular form of the density dependence of
the coupling functions $\Gamma_{m}(\varrho)$ was assumed.
The choice was motivated by results
from Dirac-Brueckner calculations of nuclear matter \cite{Typ99}. However, the
values at zero density are found from a not well constrained extrapolation to small
densities. Only near the saturation density of symmetric nuclear
matter they are well determined from model fits to the properties of
atomic nuclei. Thus, in general, one could modify the density
dependence of $\Gamma_{m}(\varrho)$ for all mesons $m$ at low densities
in order to satisfy the constraints (\ref{eq:con1}) and
(\ref{eq:con2}). Since $C_{\rho} - C_{\delta}$ has to be negative, it
is clear that the standard $\omega$, $\sigma$ and $\rho$ mesons are
not sufficient and a $\delta$ meson is needed in this case.

Since Eqs. (\ref{eq:con1}) and (\ref{eq:con2}) give only two
relations for four coupling coefficients $C_{m}$,
obtained from the consistency with the virial expansion,
two further constraints are
required to fix all couplings unambiguously.
Hence, in addition to the virial constraints, we tried to fit
the vacuum couplings coefficients $\Gamma_{m}(0)$ by describing low-energy
nucleon-nucleon scattering, i.e.\ experimentally determined NN phase
shifts \cite{Sto93}, simultaneously satisfying
equations (\ref{eq:con1}) and (\ref{eq:con2}).
The corresponding Schr\"{o}dinger equation in coordinate space was solved
using $\sigma$, $\omega$, $\rho$ and $\delta$ mesons with a
one-boson exchange potential \cite{EricsonWeise} in order
to fit the coupling constants of the nucleon-meson interaction
and to describe the experimentally known phase
shifts and effective range parameters given in \cite {Wir01},
\cite{How98}. 
The gradient terms in the potential
lead to unphysical singularities at $r\to 0$, thus to avoid these
singularities 
a certain cutoff was introduced as in \cite{Bry69}.
Coupling constants derived in this way would serve as an additional constraint
at zero densities, improving the density dependence of the coupling
functions $\Gamma_{m}(\varrho)$. However, even
when the contribution of the pion was included, that does not appear
in the mean-field approximation, it was impossible to find a
reasonable parametrization. Quite drastic modifications of
the low-density meson-nucleon couplings and meson masses were
required.
This failure is not very surprising because the number
of meson fields in the RMF model is much smaller than that in realistic
NN potentials resulting in a smaller number of degrees of freedom. In addition, the meson fields in the
RMF model do not represent real physical mesons and their masses, that
are used as adjustable parameters, e.g. in case of the $\sigma$ meson,
are not identical to the true meson masses.  
Hence, we do not follow this approach but extend the gRMF model further.

\section{Extension of the gRMF approach}
\label{sec:ExtRMF}

The failure of the gRMF model to 
describe the virial limit 
suggests that a further extension of the
approach is inevitable. In addition to the bound state two-body
correlation, the deuteron,
we introduce additional degrees of freedom in the grand
canonical potential that represent the possible two-nucleon
states, $nn$, $pp$ and $np$, in the continuum. In the following, we will
consider the isospin $0$ and $1$ channels for the $np$ states separately.
Each of the two-body channels is represented by
a temperature dependent resonance energy
($\tilde{E}_{nn}$, $\tilde{E}_{pp}$,
$\tilde{E}_{np0}$, $\tilde{E}_{np1}$)
and a corresponding effective degeneracy factor
($\tilde{g}_{nn}$, $\tilde{g}_{pp}$,
$\tilde{g}_{np0}$, $\tilde{g}_{np1}$).
In general, the resonance energies and degeneracy factors do not
necessarily have to be
the same as the resonance energies of Eq. (\ref{eq:Eres}) or the
standard degeneracy factors
in Eqs. (\ref{eq:bnn}) to (\ref{eq:bnp0}).
Two-body correlations in the continuum are considered as two-body
clusters with mass
\begin{equation}
\label{eq:mtilde}
  \tilde{m}_{ij} = N_{ij} m_{n} + Z_{ij} m_{p} + \tilde{E}_{ij}
\end{equation}
with a temperature dependent resonance energy $\tilde{E}_{ij}$.
We assume that the energy shift of a cluster $\Delta B_{ij}$
that appears in the scalar self-energy (\ref{eq:Si})
is identical to the binding energy shift of the deuteron $\Delta B_{d}$
for all continuum channels.
Of course, this most simple choice can be substituted by improved
descriptions in the future. 
The specific functional form of the 
medium-dependent energy shift will affect the form of the transition 
from very low densities, where the VEoS is applicable, to the high-density
regime. However, for the low-density matching of the VEoS and
gRMF model it is irrelevant.

The total rest mass of the cluster in the medium is given by
\begin{equation}
 \tilde{M}_{ij} = \tilde{m}_{ij} + \Delta B_{ij} \: .
\end{equation}
In case of the virial approach, the appearing masses are given by
$m_{ij} = N_{ij} m_{n} + Z_{ij} m_{p} + E_{ij}$ and
$M_{ij} = m_{ij}$, respectively.
We include the effect of the resonance energies in
the relativistic correction through the
$k_{2}$ function
and the definition of the thermal wave lengths
$\lambda_{ij}=[2\pi/(M_{ij}T)]^{1/2}$  and
similarly for $\tilde{\lambda}_{ij}$.
With these modifications we can rederive consistency
relations between this extended gRMF model and the VEoS.

The virial integrals (\ref{eq:virint}) in the virial coefficients
are represented by an exponential of the effective resonance
energy as defined in (\ref{eq:Eres}). The two $np$ channels are
separated by requiring for the $np$ isospin 1 channel a similar
form as for the $nn$ and $pp$ channels. Comparing the fugacity
expansion of the gRMF model with 2-body correlations with the VEoS
at low densities, we finally obtain an extension of the relations
(\ref{eq:nn_comp}) to (\ref{eq:np_comp}) , where an additional term
with temperature dependent resonance energy $\tilde{E}_{ij}(T)$
appears in the right-hand side of the Eqs.
(\ref{eq:nn_2}) to (\ref{eq:np0_2}), i.e.\ 
\begin{eqnarray}
\label{eq:nn_2}
 \lefteqn{-\frac{T}{\lambda_{ij}^{3}}
 k_{2}\left( \frac{M_{ij}}{T}\right)
 \hat{g}_{ij} \exp \left( - \frac{E_{ij}}{T} \right)}
 \\ \nonumber & = &
  -\frac{T}{\tilde{\lambda}_{ij}^{3}}
 k_{2}\left( \frac{\tilde{M}_{ij}}{T}\right)
 g^{\rm (eff)}_{ij} \exp \left( - \frac{\tilde{E}_{ij}}{T} \right)
 \\ \nonumber & &
 + \frac{1}{2} \frac{g_{i}g_{j}}{\lambda_{i}^{3}\lambda_{j}^{3}}
 \left[ \left( C_{\omega}  + C_{\rho} \right)
   k_{2}\left( \frac{m_{i}}{T}\right)  k_{2}\left(
     \frac{m_{j}}{T}\right)
 \right. \\ \nonumber & & \left.
 -  \left( C_{\sigma}  + C_{\delta} \right)
  k_{1}\left( \frac{m_{i}}{T}\right) k_{1}\left( \frac{m_{j}}{T}\right)
 \right]
\end{eqnarray}
for the channels $ij= nn$, $pp$, $np1$ and
\begin{eqnarray}
\label{eq:np0_2}
 \lefteqn{-\frac{T}{\lambda_{np0}^{3}}
 k_{2}\left( \frac{M_{np0}}{T}\right)
 \hat{g}_{np0} \exp \left( - \frac{E_{np0}}{T} \right)}
 \\ \nonumber & = &
  -\frac{T}{\tilde{\lambda}_{np0}^{3}}
 k_{2}\left( \frac{\tilde{M}_{np0}}{T}\right)
 g^{\rm (eff)}_{np0} \exp \left( - \frac{\tilde{E}_{np0}}{T} \right)
 \\ \nonumber & &
 + \frac{1}{2} \frac{g_{n}g_{p}}{\lambda_{n}^{3}\lambda_{p}^{3}}
 \left[ \left( C_{\omega}  - 3C_{\rho} \right)
 k_{2}\left( \frac{m_{n}}{T}\right) k_{2}\left( \frac{m_{p}}{T}\right)
 \right. \\ \nonumber & & \left.
 - \left( C_{\sigma}  - 3C_{\delta} \right)
 k_{1}\left( \frac{m_{n}}{T}\right) k_{1}\left( \frac{m_{p}}{T}\right)
 \right]
\end{eqnarray}
for the channel $np0$.
These consistency relations leave sufficient freedom to reproduce the
exact VEoS at low densities in the gRMF model. There are
different possibilities to achieve this goal depending
on the choice of the zero-density couplings and on the choice
of the effective resonance energies and degeneracy factors that
represent the continuum contributions in the extended gRMF model.
The structure on the left-hand side shares some similarities
with the generalized Beth-Uhlenbeck approach in Section \ref{sec:GBU}.
One part of the effects of the two-body correlations is included in
the explicit continuum contributions
and the remaining part is covered by the mean-field terms.
There is the freedom to shift the correlation
strength between these terms rather arbitrarily. In the following, a
few choices will be discussed.

In Section \ref{sec:T0limit} the case without introducing explicit
continuum channels in the gRMF model, i.e.\ shifting the full
correlation strength into a redefinition of the vacuum couplings, 
turned out to be impossible.
The other extreme is to represent the
correlations by the introduction of the effective continuum states
described by temperature dependent
resonance energies $\tilde{E}_{ij}$ with
medium corrections $\Delta B_{ij}$ as described above.
In this case the contribution of the virial integral in the left-hand
side of the Eqs. (\ref{eq:nn_2}) and (\ref{eq:np0_2})
is fully represented by the the resonance energies $\tilde{E}_{ij}$ 
in the first term in the right-hand side of these equations.
Therefore the mean-field part has to vanish in the non-relativistic
limit, leading to
$  C_{\omega} - C_{\sigma} = 0$
and
$  C_{\rho} - C_{\delta} = 0 $.
The coupling coefficients $C_m$ in relations
(\ref{eq:con1}) to (\ref{eq:con2}) were expressed through the virial
integral taken in the effective range approximation. In the new
relations, given above, the virial integral and thus the
scattering lengths are implemented in the definition of the
resonance energies $\tilde{E}_{ij}(T)$. Therefore, neglecting the
term proportional to $ g^{\rm (eff)}_{ij} \exp \left( -
\tilde{E}_{ij}/T \right)$ in the relations
(\ref{eq:nn_2}) and (\ref{eq:np0_2}), we recover
equations (\ref{eq:nn_comp}) to (\ref{eq:np_comp}), leading to the
consistency conditions (\ref{eq:con1}) to (\ref{eq:con2}).

In order to fix the four couplings $C_m$ unambiguously, we again
tried to describe low-energy nucleon-nucleon scattering, by
solving the Schr\"{o}dinger equation with one-boson exchange
potentials as described in Section \ref{sec:T0limit} \,
simultaneously with the two conditions on the couplings $
C_{\omega} - C_{\sigma} = 0$, $  C_{\rho} - C_{\delta} = 0$.
We observed that the required coupling constants at zero density are drastically
different in comparison with the previous values of the
parametrization DD2 or DD-ME$\delta$
leading to problems such as a rather unphysical behavior of the EoS at
low densities by similar reasons as in the first attempt 
at the end of Section \ref{sec:T0limit}. 
Consequently, this approach was discarded. 
It was not possible to describe both the NN scattering data and the 
VEoS simultaneously 
in this second choice for representing the continuum correlations.

Therefore the most straightforward way to fulfill the consistency
relations  (\ref{eq:nn_2}) and (\ref{eq:np0_2}) 
is to leave the density dependence of meson-nucleon couplings untouched as given
by the gRMF model parametrization and to turn down the idea of an additional
fit to NN scattering data, thus concentrating only on the
reproduction of the VEoS. 
The effects of the two-body correlations are incorporated in the
effective resonance energies that are chosen as identical to those of the VEoS, i.e.\
$\tilde{E}_{ij} = E_{ij}$.
Then, the consistency relations
(\ref{eq:nn_2}) and (\ref{eq:np0_2}) are seen as the defining equations
for the effective degeneracy factors $g^{\rm (eff)}_{ij}$ that now
depend on temperature. 
In the non-relativistic limit, these equations
simplify as
\begin{eqnarray}
\label{eq:1S0nn}
 g^{\rm (eff)}_{nn}(T)
 & = &
   \hat{g}_{nn}
 +  \frac{g_{n}^{2}}{2T} \frac{\lambda_{nn}^{3}}{\lambda_{n}^{6}}
 C_{1} \exp \left[ \frac{E_{nn}(T)}{T} \right]  \: ,
 \\
 g^{\rm (eff)}_{pp}(T)
 & = &
 \hat{g}_{pp}
 + \frac{g_{p}^{2}}{2T} \frac{\lambda_{pp}^{3}}{\lambda_{p}^{6}}
 C_{1} \exp \left[  \frac{E_{pp}(T)}{T} \right] \: ,
 \\
 g^{\rm (eff)}_{np1}(T)
 & = &
 \hat{g}_{np1}
 +  \frac{g_{n}g_{p}}{2T} \frac{\lambda_{np1}^{3}}{\lambda_{n}^{3}\lambda_{p}^{3}}
 C_{1} \exp \left[ \frac{E_{np1}(T)}{T} \right]
  \: ,
 \\
 \label{eq:3S1np}
 g^{\rm (eff)}_{np0}(T)
 & = &
 \hat{g}_{np0}
 + \frac{g_{n}g_{p}}{2T} \frac{\lambda_{np0}^{3}}{\lambda_{n}^{3}\lambda_{p}^{3}}
  C_{0} \exp \left[ \frac{E_{np0}(T)}{T} \right]
\end{eqnarray}
where
\begin{eqnarray}
 C_{1} & = &  C_{\omega} -  C_{\sigma} + C_{\rho}  - C_{\delta}  \: ,
 \\
 C_{0} & = &  C_{\omega} -  C_{\sigma}  - 3\left( C_{\rho} -
   C_{\delta} \right)
\end{eqnarray}
are the relevant couplings in the two isospin channels. If
$C_{1}=C_{0}=0$ as was discussed before as a possible condition, 
we recover the standard
degeneracy factors $\hat{g}_{ij}$ in the s-wave. The degeneracy
factors on the right-hand side are $ \hat{g}_{nn} = \hat{g}_{pp} =
\hat{g}_{np1} = 1 $ and $ \hat{g}_{np0} = -3 $, respectively. The
negative sign for $ \hat{g}_{np0}$ is related to the positive
s-wave scattering length in this channel. Note that in the NSE
description, cf.\ Subsection \ref{sec:NSE}, the temperature
dependence of the degeneracy factors follows from the inclusion of
excited states of the nuclei. In our approach, it represents the
contribution of the continuum states and thus is very similar.

\begin{figure}[t]
  \begin{center}
    \includegraphics[width=0.7\linewidth]{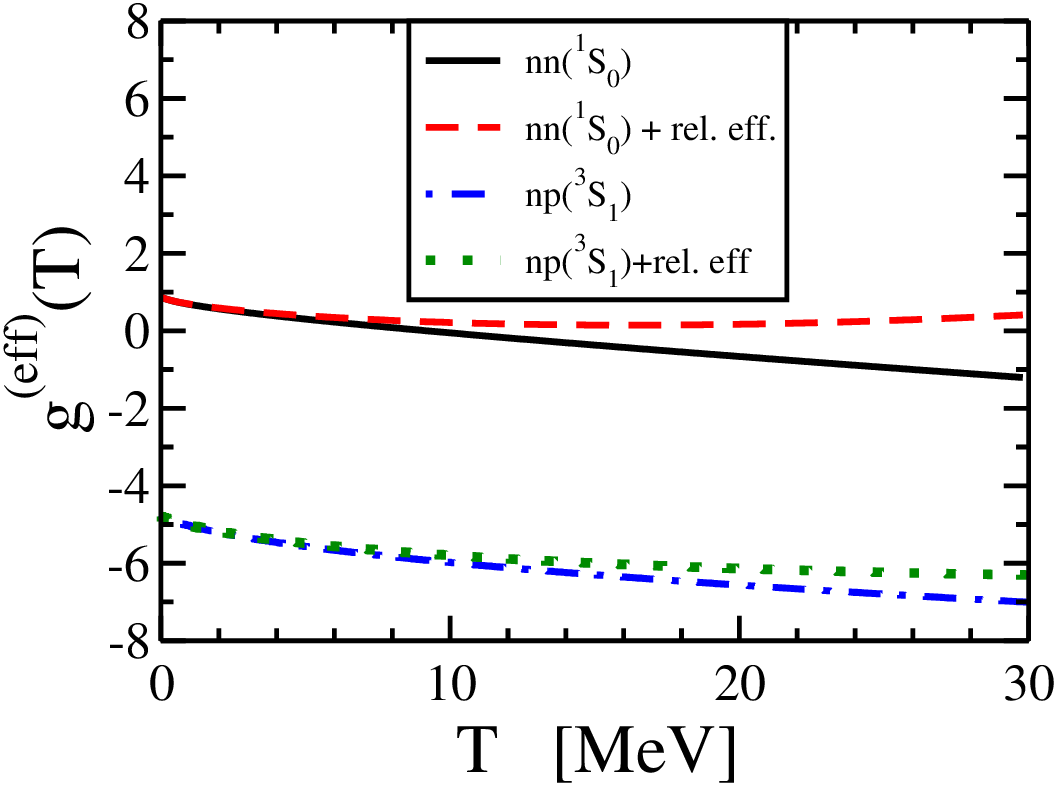}
    \caption{\label{fig:degfact}%
      Temperature dependence of the effective degeneracy
      factors $g^{\rm (eff)}_{nn}$ and $g^{\rm (eff)}_{np0}$ with and without
      relativistic corrections.}
  \end{center}
\end{figure}

In Fig.\ \ref{fig:degfact} the temperature dependence of the effective
degeneracy factors is depicted for the scattering correlations in the $nn$ and
$np0$ channels. Because the curves for the $pp$ and
$np1$ channels are practically identical to the
$nn$ scattering case, they are not shown.
The fully
relativistic results as they follow from Eqs. (\ref{eq:nn_2}) and
(\ref{eq:np0_2}) are compared with their non-relativistic limits
(\ref{eq:1S0nn}) and (\ref{eq:3S1np}),
respectively. We see that the relativistic corrections 
caused by the $k_{1}$ and $k_{2}$ functions are
minor at low temperatures but grow
with the increase of temperature as expected, being
larger for the case of $nn$ scattering. The large relativistic
correction in the effective degeneracy factors
is caused by an accidental change of sign in the factor
containing the coupling coefficients in Eq. (\ref{eq:nn_2}).

In the zero-temperature limit, the effective degeneracy factors
are given by
\begin{eqnarray}
\label{eq:1S0nn_b}
 g^{\rm (eff)}_{ij}(0)
 & = &
   \hat{g}_{ij}
 -  \frac{\hat{g}_{ij}}{g_{ij}} \frac{C_{1}}{2\pi a_{ij}} 
 \frac{2m_{i}m_{j}}{m_{i}+m_{j}} \: ,
 \\
 \label{eq:3S1np_b}
 g^{\rm (eff)}_{np0}(0)
 & = &
 \hat{g}_{np0}
 -  \frac{\hat{g}_{np0}}{g_{np0}} \frac{C_{0}}{2\pi a_{np0}} 
 \frac{2m_{n}m_{p}}{m_{n}+m_{p}}
\end{eqnarray}
for the channels $ij=nn,pp,np1$ and $np0$, respectively, when the
approximation (\ref{eq:I0a}) with the scattering length is used.
Thus, there is still a difference as compared to $\hat{g}_{ij} = 1$
and $\hat{g}_{np0} = -3$, respectively, when the correction due to the
meson couplings is not taken into account. This is a clear
indication that the mean field effectively takes over a part
of the correlations.

%

\section{Neutron matter}
\label{sec:neumat}

In nuclear matter with arbitrary neutron to proton ratio, the two-body
correlations in all nucleon-nucleon channels contribute to the
thermodynamic properties and, in particular, the deuteron bound state
dominates at low temperatures. The case is different for pure neutron
matter where only the $nn$ channel without a bound state is
relevant. Thus, neutron matter is ideally suited to demonstrate the
effects of the continuum correlations in the extension of the
gRMF model. Before the general case of finite temperatures
is discussed, we first consider in brief the 
case $T=0$ in the limit of small densities.

\subsection{Zero temperature limit}
\label{sec:zeroT}

At zero temperature the Fermi momentum $k_{F_{n}}$ defined
through the neutron density
$ n_{n} = k_{F_{n}}^{3}/(3\pi^{2})$
sets the scale for all results. Instead of an expansion in
powers of the neutron fugacity at finite temperature, a series expansion in powers of
$k_{F_{n}}$ is the relevant method to study the low-density behavior of
the EoS.
In Ref.\ \cite{Lee57} it was shown that
the energy per neutron $E/N$ (without rest mass) at very small $k_{F_{n}}$ can be
expanded in a power series
\begin{equation}
\label{eq:EN}
 \frac{E}{N} = 
 E_{\rm free} \xi
\end{equation}
with the power series
\begin{equation}
\label{eq:LeeYang}
 \xi = 
 1 + \frac{10}{9\pi} \zeta 
 + \frac{4}{21\pi^{2}} \left( 11 - 2 \ln 2\right)
 \zeta^{2} 
 + \dots
\end{equation}
in the dimensionless parameter $\zeta$ = $a_{{}^{1}S_{0}}^{(nn)}k_{F_{n}}$.
Here $E_{\rm free}=3k_{F_{n}}^{2}/(10m_{n})$ is the energy per neutron
of a non-interacting Fermi gas that defines the relevant energy scale.
Because of
the unnaturally large negative nn scattering length $a_{{}^{1}S_{0}}^{(nn)}\approx
-18.8$~fm \cite{Wir01}, the radius of convergence is unfortunately very small,
i.e.\ $n_{n} \ll 243\pi/|10a_{{}^{1}S_{0}}^{(nn)}|^{3} \approx 1.1 \cdot 10^{-4}$~fm$^{-3}$.

In the DD-RMF approach without effective contributions of two-body
correlations, an expansion of the energy per neutron in powers of the
Fermi momentum (see \ref{sec:RMFlow}) leads to
\begin{equation}
\label{eq:ENRMF}
 \frac{E}{N} =E_{\rm free} \xi_{\rm RMF} 
\end{equation}
with
\begin{equation}
\label{eq:xiRMF}
 \xi_{\rm RMF} = 
 1 + \frac{5}{9\pi^{2}}
 \left(  C_{\omega} -  C_{\sigma} + C_{\rho}  - C_{\delta} \right)
 m_{n} k_{F_{n}} + \dots
 \: .
\end{equation}
A comparison with Eq.\ (\ref{eq:LeeYang}) gives the relation
\begin{equation}
  C_{\omega} -  C_{\sigma} + C_{\rho}  - C_{\delta}
 = \frac{2\pi}{m_{n}}
 a_{{}^{1}S_{0}}^{(nn)}.
\end{equation}
This condition coincides with condition (\ref{eq:nn_comp})
for neutron matter,
although it is derived in the low 
$k_{F_{n}}/m_{n}$
limit for $T=0$
whereas condition (\ref{eq:nn_comp}) in the $T \to 0$ limit of the finite-temperature
virial expansion with vanishing convergence radius in density.
From a physical point of view, this coincidence is not surprising
since both approaches only use two-body scattering information.
However, it is gratifying to find the coincidence from the two very
different approaches.
We find the values
of $-2.91$~fm$^{2}$ and $-2.35$~fm$^{2}$ for
$C_{\omega} -  C_{\sigma} + C_{\rho}  - C_{\delta}$ in the DD2 and
DD-ME$\delta$ parametrizations, respectively. However, these are much
smaller in modulus as compared to the required
value of $2\pi a_{{}^{1}S_{0}}^{(nn)}/m_{n} = -24.83$~fm$^{2}$.

A particular situation arises in the unitary limit \cite{Dru11},
i.e.\ if $a_{{}^{1}S_{0}}^{(nn)}$ approaches $-\infty$.
Then the series expansion
(\ref{eq:LeeYang}) can no longer be applied since the radius of
convergence shrinks to zero. In fact, the energy per
nucleon should scale as
the energy per neutron
of a non-interacting Fermi gas $E_{\rm free}$
with a universal constant $\xi$ independent of $k_{F_{n}}$. 
The parameter $\xi_{\rm RMF}$ is
independent of $k_{F_{n}}$ in first-order only if
the combination 
$C_{\omega} -  C_{\sigma} + C_{\rho}  - C_{\delta}$ diverges as
$k_{F_n}^{-1}$. Hence a
particular density dependence of at least one meson-nucleon coupling
is required in this case.

In the nonlinear RMF model of Ref.\ \cite{GShe10a} with nonlinear
self-interactions of the scalar meson, a density dependence of the $\sigma$ meson
coupling was introduced for densities lower than
a particularly chosen transition density with the aim to reproduce the energy per
neutron for unitary neutron matter assuming $\xi = 0.44$ \cite{Car03}. This
condition required a divergence of $\Gamma_{\sigma}(n) \propto n^{-1/6}$
that is consistent with the expectation from Eq.\
(\ref{eq:xiRMF}). However, mixing the nonlinear RMF approach
and the RMF model with density dependent and divergent couplings does not seem to
be very natural. In our approach, we do not aim to describe the neutron
matter EoS at zero temperature and low densities as a unitary Fermi
gas (FG) but require consistency with the VEoS for $T>0$.

\subsection{Finite temperatures}

We choose $T=4$~MeV and $T=10$~MeV as representative examples
which illustrate the main effects due to interactions and correlations. 
Different approximations in the description will be
compared, ranging from the ideal
Fermi gas via the virial
approach to the gRMF model. Nucleons are treated as Fermions
and clusters as Maxwell-Boltzmann particles
because they appear at high temperatures with low abundancies
when deviations from the correct Bose-Einstein statistics are negligible.

There are two quantities that exhibit finite
low-density limits and
hence are very advantageous for a comparison of the models:
the ratio of the pressure over
total particle number density $p/n$ and of the internal energy per baryon $E/N
= \varepsilon/n-m_{n}$ (without the rest mass of the neutron) with the energy density
\begin{equation}
\label{eq:epsilon}
 \varepsilon = \frac{1}{V} \left(\Omega + T S \right)
 + \sum_{i=n,nn} \tilde{\mu_{i}} n_{i}
\end{equation}
that can be calculated from the
grand canonical potential
$\Omega$, the entropy $S$ and the relativistic chemical potentials $\tilde{\mu}_{i}$.
The sum contains the contribution of the neutrons and that of the
correlated two-neutron continuum. Correspondingly,
the total particle number density $n$ is the sum
\begin{equation}
 n = n_{n} + 2 n_{nn}
\end{equation}
and the mass fraction of correlated two-neutron states is defined by
\begin{equation}
 X_{nn} = \frac{2n_{nn}}{n_{n}+2n_{nn}} \: .
\end{equation}
For an ideal
gas with Maxwell-Boltzmann statistics and non-relativistic kinematics,
we have the simple result that $p/n = T$ and $E/N = 3T/2$ are independent
of the density of the system and trivially $X_{nn}=0$.
Deviations from these values indicate
the effects of correlations and interactions.

\subsubsection{Low densities}
\label{subsec:nm_low}

\begin{figure}[t]
  \begin{center}
    \epsfig{file=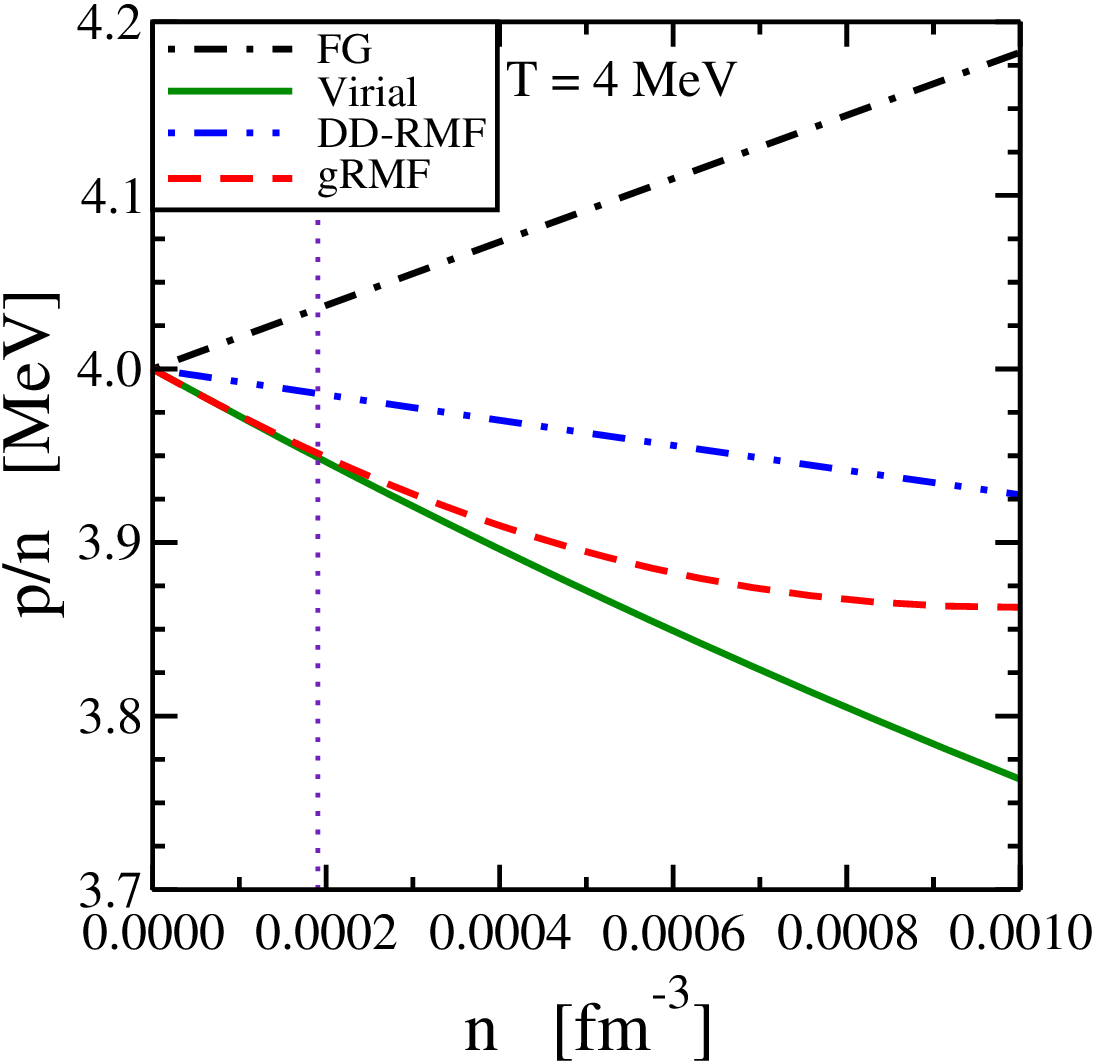,width=0.47\textwidth}
    \epsfig{file=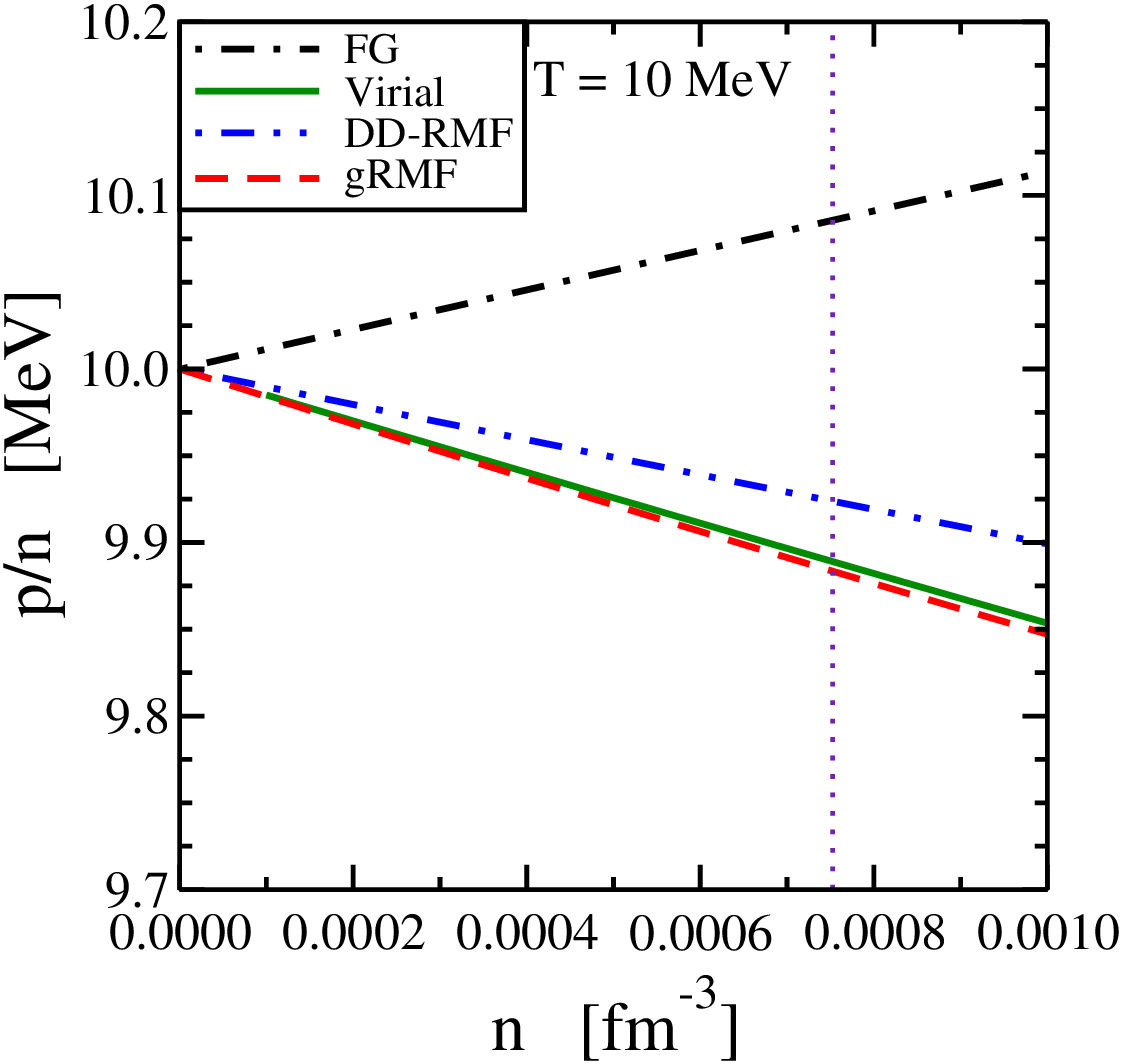,width=0.48\textwidth}
    \caption{\label{fig:pn}%
      Ratio of pressure over total particle number density, $p/n$,
      of neutron matter as a function of the total density $n$
      for temperatures of $T=4$~MeV (left) and $T=10$~MeV (right).
      Vertical dotted lines indicate the density where $n\lambda_{n}^{3}=1/10$.}
  \end{center}
\end{figure}

\begin{figure}
  \begin{center}
    \epsfig{file=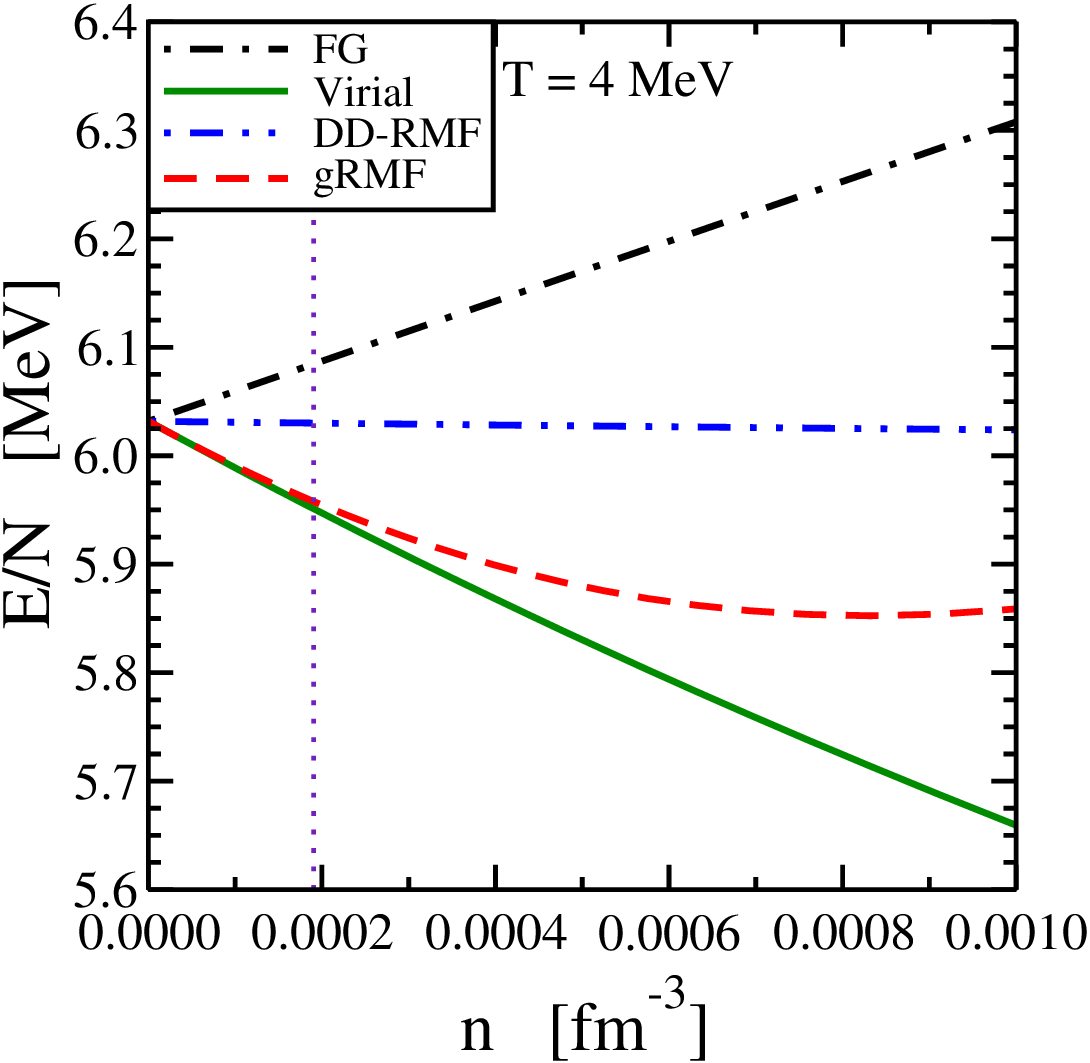,width=0.47\textwidth}
    \epsfig{file=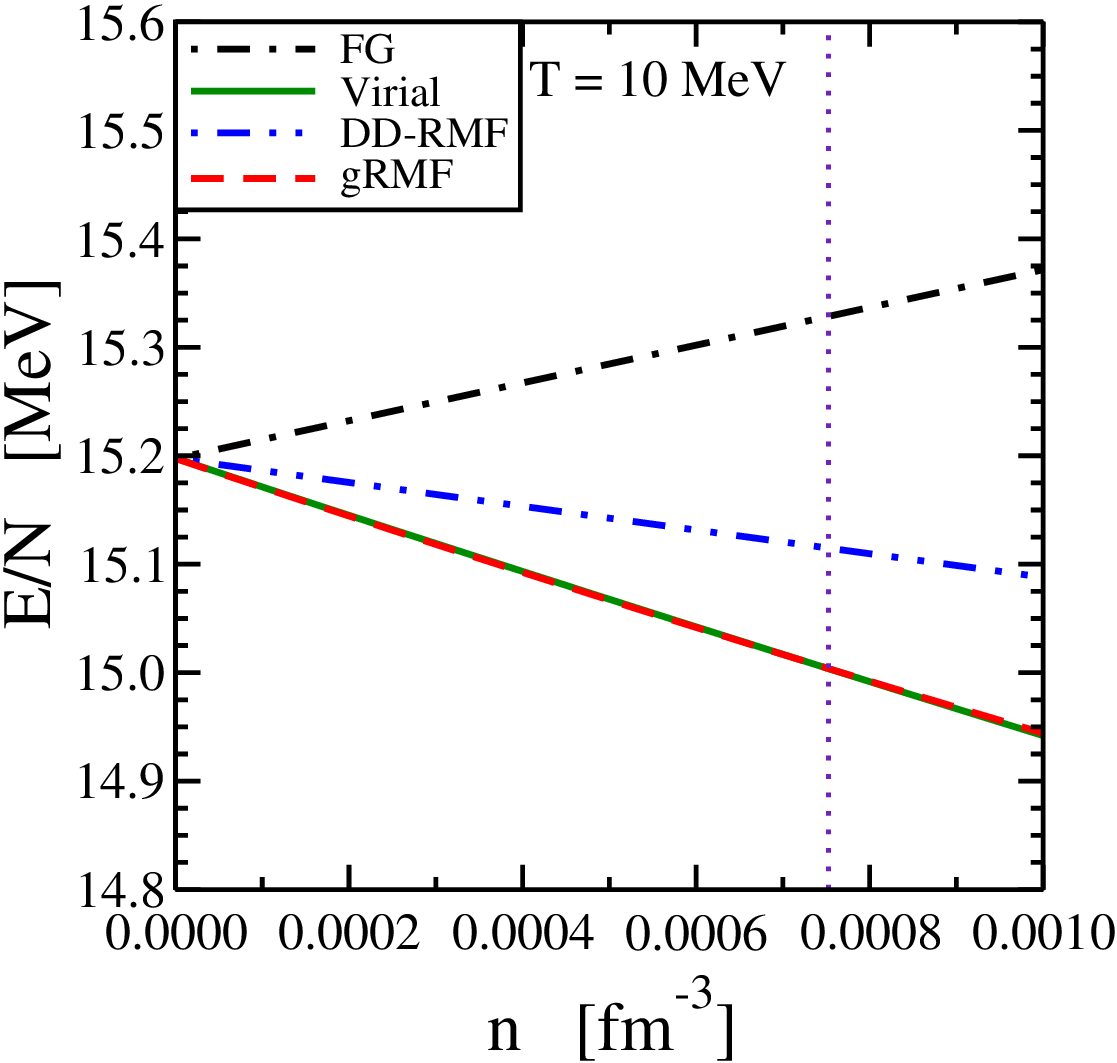,width=0.48\textwidth}
    \caption{\label{fig:EN}%
      Internal energy per baryon (without contribution of the neutron rest mass)
      in neutron matter as a function of the total particle number
      density $n$ for temperatures of $T=4$~MeV (left) and $T=10$~MeV
      (right).
      Vertical dotted lines indicate the density where $n\lambda_{n}^{3}=1/10$.}
  \end{center}
\end{figure}

In Figures
\ref{fig:pn} and \ref{fig:EN} the two quantities $p/n$ and $E/N$,
respectively, are depicted for the
two selected temperatures as a function of the total particle number
density $n$ in different theoretical approaches.
In case of the relativistic Fermi gas (see FG-curve),
effects of the Pauli principle are taken into account leading to an increase of the
pressure and of the energy per neutron as compared to the ideal Boltzmann gas. The
limit $\lim_{n\to 0} (p/n) = T$ is not affected by statistical
corrections or
relativistic kinematics since the $k_{2}$ factors cancel in the
lowest order of the fugacity expansion. This is easily seen considering the ratio
$p=-\Omega/(Vn)$ using Eqs.\ (\ref{eq:Omegai_2}) and (\ref{eq:ni_2}).
In contrast, the relativistic correction factor $k_{2}$ in
Eq.\ (\ref{eq:ni_2}) modifies the relation between the neutron density and
neutron chemical potential appearing in Eq.\ (\ref{eq:epsilon}) and
thus $\lim_{n \to 0} (E/N)> 3T/2$ in figure \ref{fig:EN}. 
The relativistic corrections become
larger with increasing temperature. The shift of $E/N$ at zero
density can be estimated as $2T^{2}/m_{n}$.

The VEoS predicts a dependence of $p/n$ and $E/N$ on the density
with a negative slope. This is the effect of the correlations induced
by the $nn$ interaction. The vertical lines in Figures \ref{fig:pn}
and \ref{fig:EN} denote the density $n$ where $n \lambda_{n}^{3} = 1/10$. At
densities above this value, higher-order contributions to the VEoS,
that are not considered in the fugacity expansion up to second order, can
be expected to contribute significantly. Figure \ref{fig:EN} also
demonstrates that the virial corrections at low densities are smaller
than the relativistic correction that leads to an overall shift.

\begin{figure}[ht]
  \begin{center}
    \includegraphics[width=0.47\linewidth]{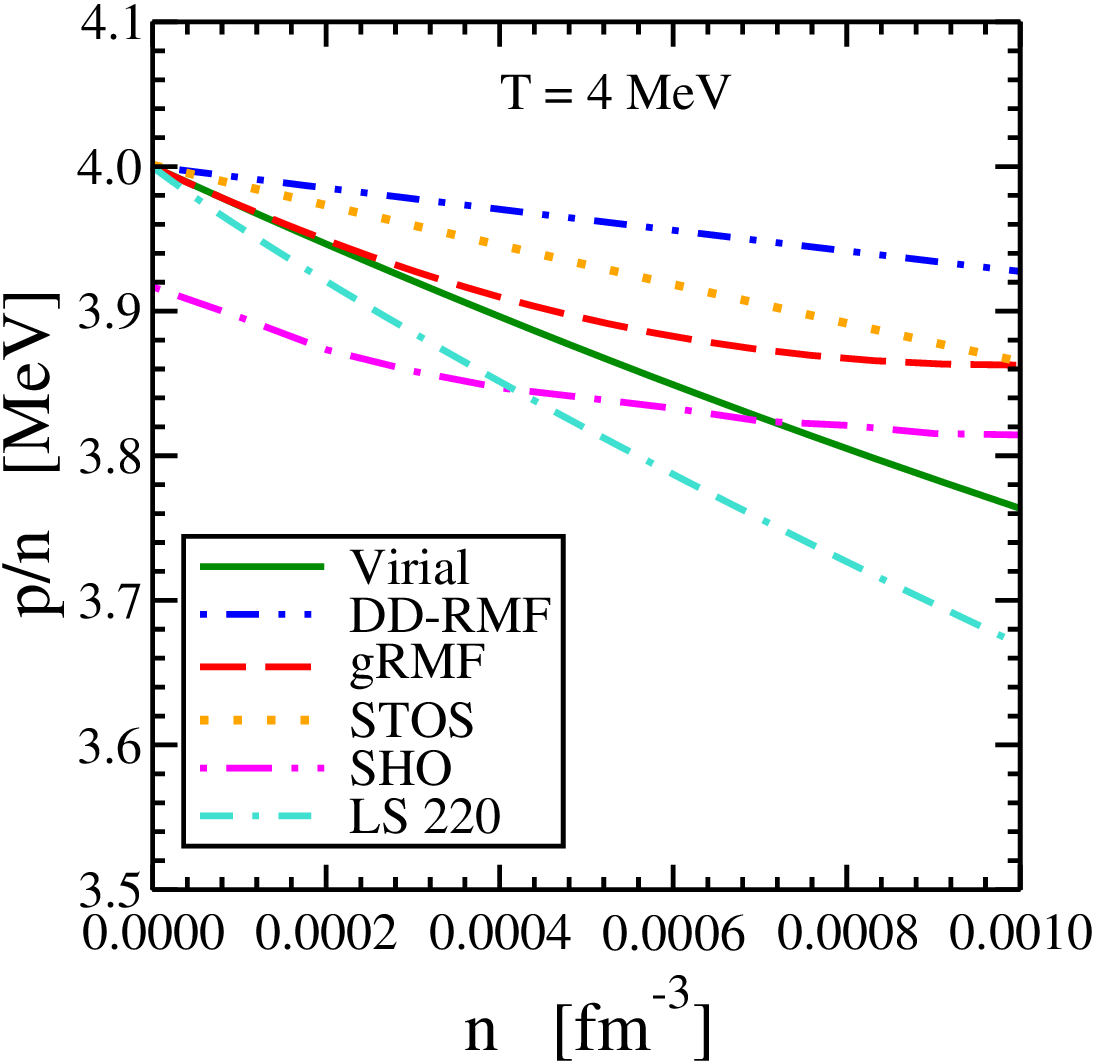}
    \includegraphics[width=0.482\linewidth]{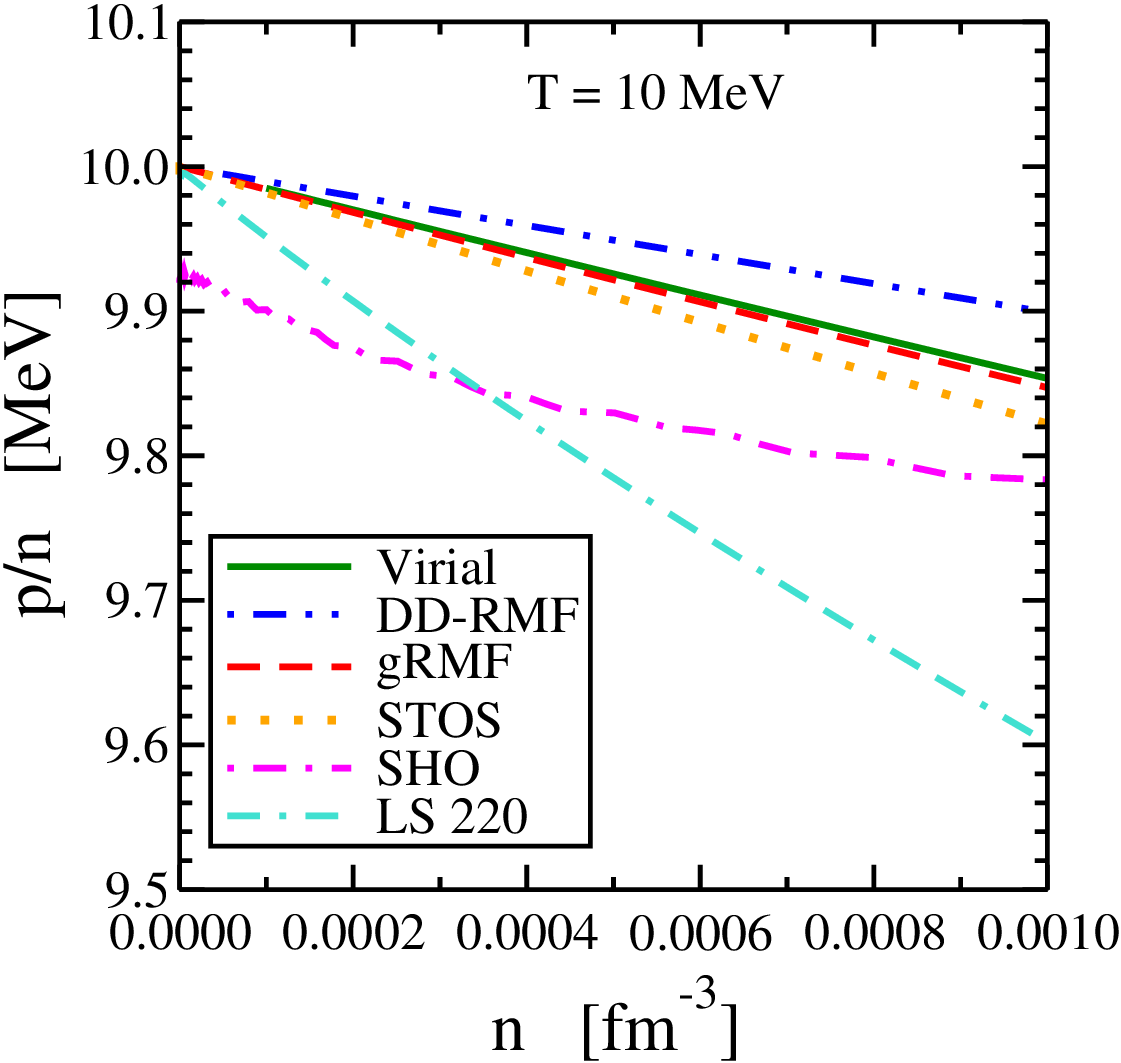}
    \caption{\label{fig:CompEoS}%
      Ratio of pressure over total particle number density, $p/n$,
      of neutron matter as a function of the density $n$
      for temperatures of $T=4$~MeV (left) and $T=10$~MeV (right)
      in different models. See text for details.}
  \end{center}
\end{figure}

The curves of the DD-RMF model with DD2 parametrization without correlations
lie between the VEoS and the FG results. They do not show
the correct dependence given by the VEoS at low densities. When the $nn$
correlations are taken into account in the gRMF model with the
quadratic form of the energy shift (see next subsection) the low-density
behavior of the VEoS is nicely reproduced. Only at higher densities,
medium effects, that are not incorporated in the  VEoS, lead to
a deviation. The precise density dependence of the deviation will
depend on the choice of the functional form of the energy shift, but
the agreement in the low-density limit is not affected.
Obviously, deviations of the gRMF EoS from the VEoS
start to become more important with increasing
density at lower temperatures.

It is also worthwhile to compare the predictions of the VEOS, the
original DD-RMF
and gRMF models in the DD2 parametrization
with the results of other approaches used in astrophysical
applications. We examine two other RMF models that employ nonlinear
self-interactions of the mesons: the model
of Shen, Horowitz and O'Connor\ (SHO) \cite{GShe11b} with the FSUGold
parametrization and the model of Shen, Toki, Oyamatsu and Sumiyoshi\ (STOS)
\cite{HShe11} with the parameter set TM1.
In addition, the non-relativistic Lattimer-Swesty EoS
with incompressibility $K = 220$~MeV (LS 220) is considered.

In Figure \ref{fig:CompEoS} the density dependence of
the quantity $p/n$ at $T=4$~MeV and $T=10$~MeV is depicted for all models below
$0.001$~fm$^{-3}$. The curves of the STOS model, that does not include
two-body $nn$ correlations, are surprisingly close to the exact VEoS
line for $T=10$~MeV,
but do not reproduce it exactly. The deviations are larger at
$T=4$~MeV and the close agreement seems to be accidental
for $T=10$~MeV. The SHO approach with the FSUGold parametrization
claims to be constructed such that the case of unitary neutron matter
is reproduced at low densities by introducing a particular density
dependent coupling of the $\sigma$ meson, see the discussion at the
end of Subsection \ref{sec:zeroT}. A large deviation from
the VEoS result can be seen in Fig.\ \ref{fig:CompEoS}
and the correct low-density limit for $p/n$ is not
reproduced. Furthermore, the tabulated data exhibit some oscillations
that are probably related to the choice of the interpolation
procedure \cite{GShe11}.
The LS EoS shows a much larger negative slope of
$p/n$ as a function of $n$ as compared to the other models but
reproduces the correct ideal gas limit.

\begin{figure}[t]
  \begin{center}
    \epsfig{file=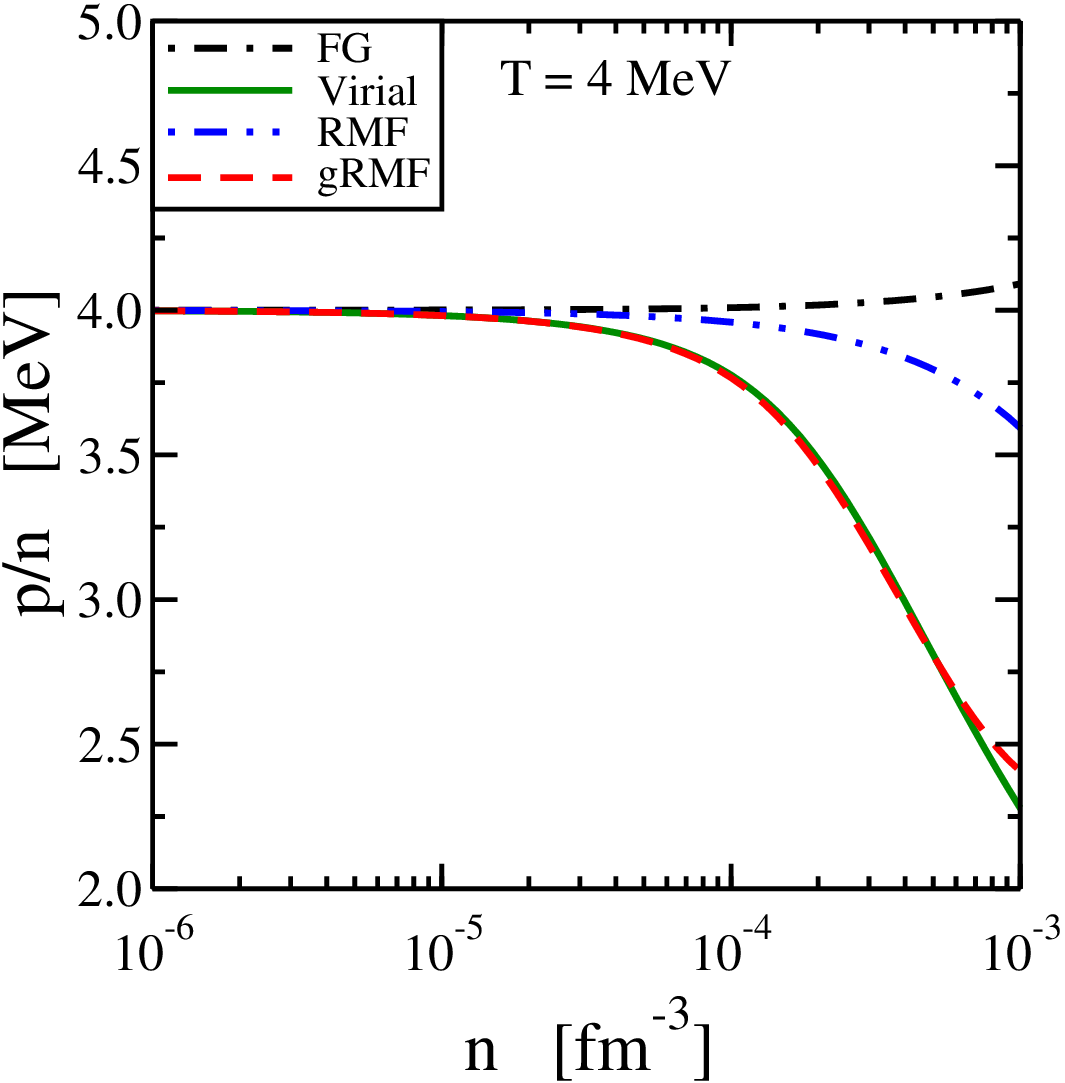,width=0.47\textwidth}
    \epsfig{file=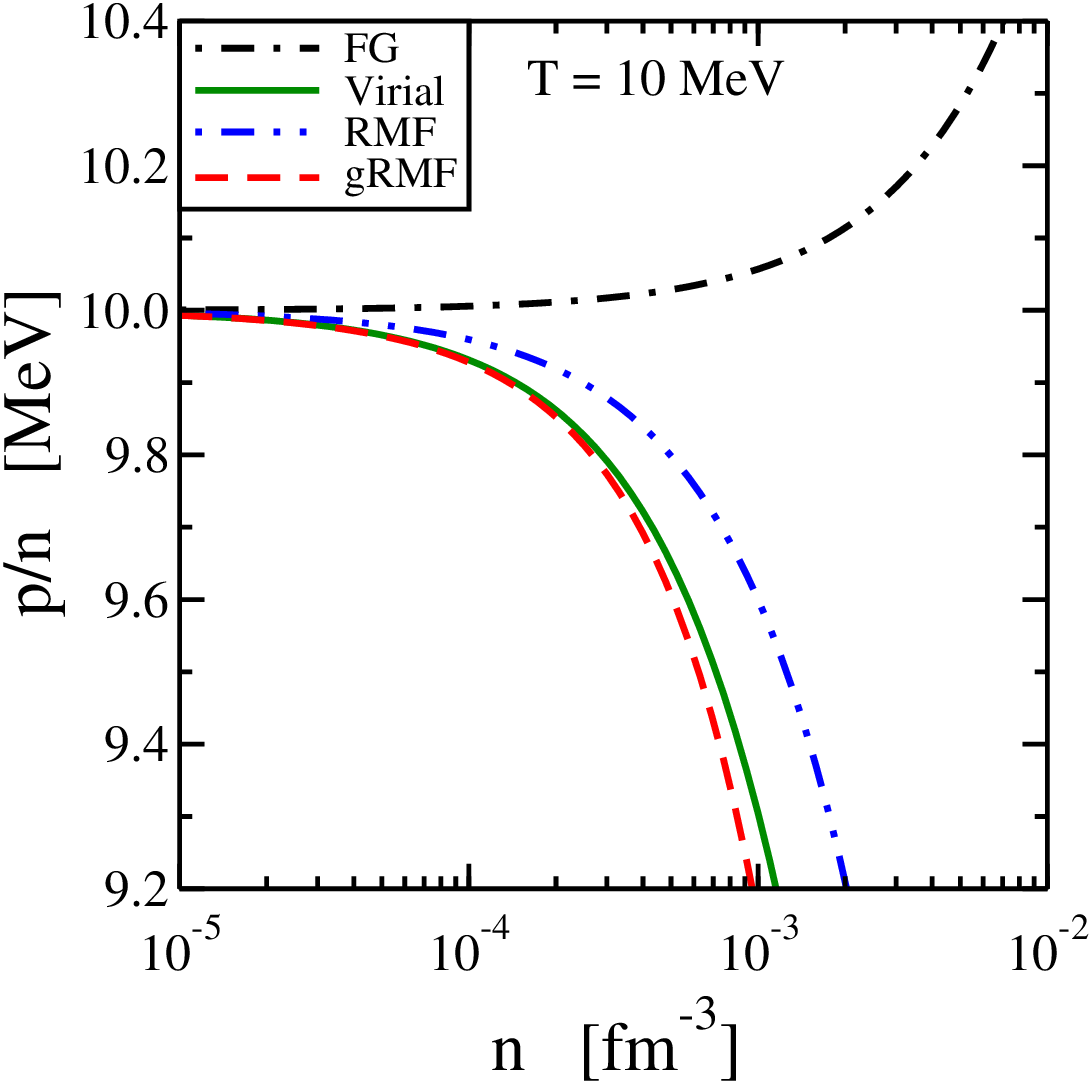,width=0.48\textwidth}
    \caption{\label{fig:psym}%
      Ratio of pressure over total particle number density, $p/n$,
      of symmetric nuclear matter as a function of the total density $n$
      for temperatures of $T=4$~MeV (left) and $T=10$~MeV (right).}
  \end{center}
\end{figure}

\begin{figure}
  \begin{center}
    \epsfig{file=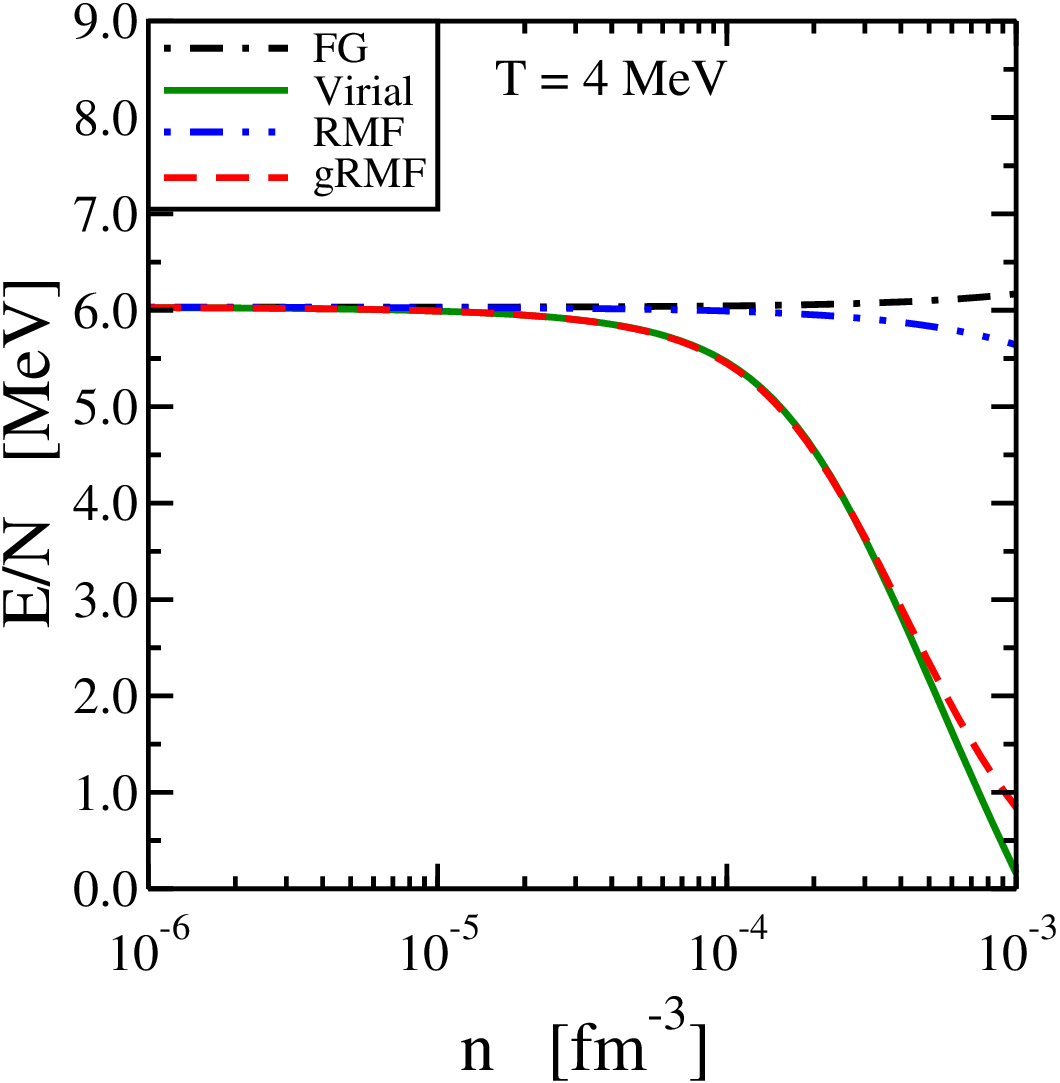,width=0.47\textwidth}
    \epsfig{file=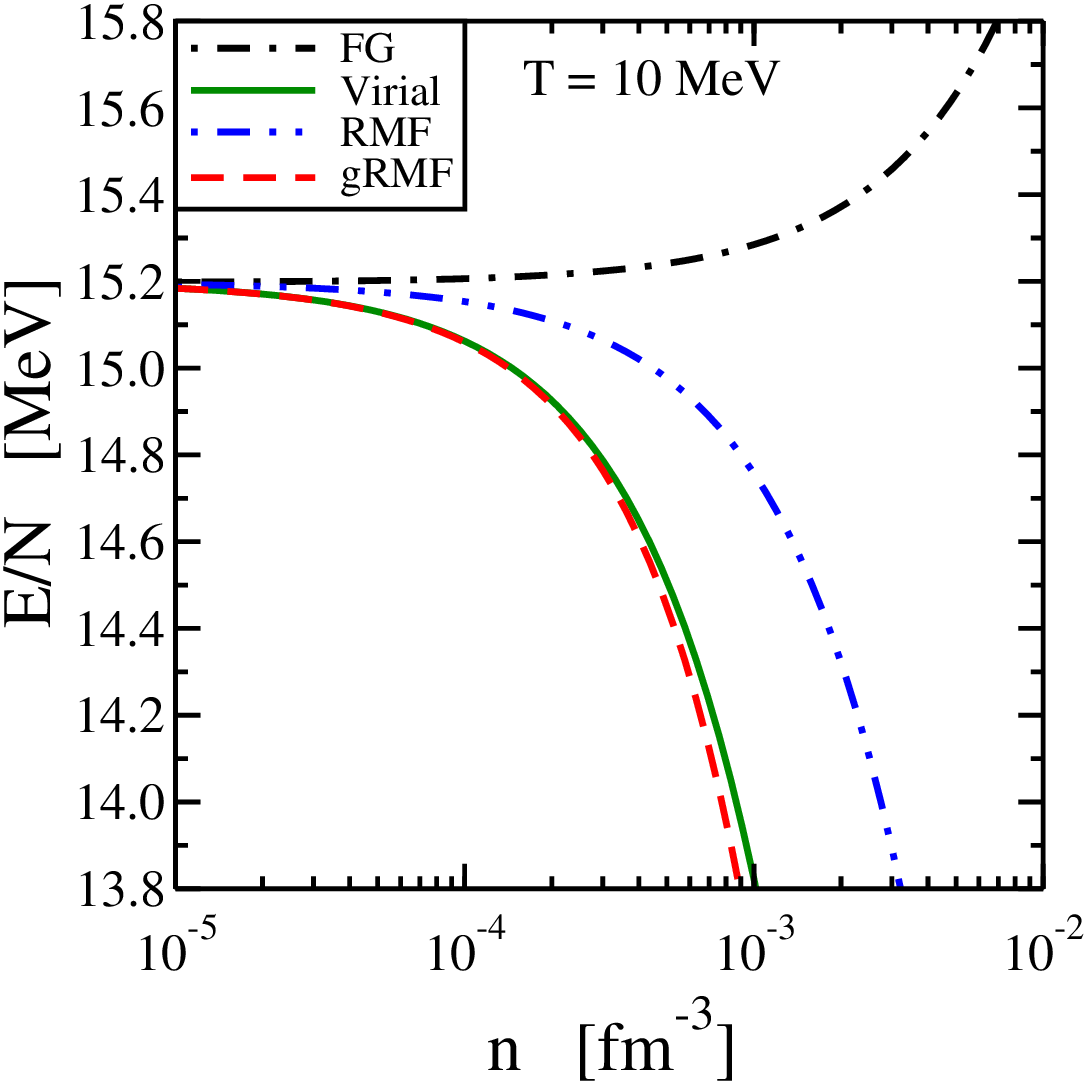,width=0.48\textwidth}
    \caption{\label{fig:Esym}%
      Internal energy per baryon (without contribution of the neutron rest mass)
      in symmetric nuclear matter as a function of the total particle number
      density $n$ for temperatures of $T=4$~MeV (left) and $T=10$~MeV
      (right).}
  \end{center}
\end{figure}

\section{Symmetric nuclear matter}
\label{sec:snm}

In contrast to neutron matter, where only neutron scattering
  correlations contribute to the thermodynamical quantities, 
  all correlations of neutrons and protons
  in scattering and bound states should 
  be taken into account in symmetric matter. 
  At low densities, two-body correlations will
  be most important, but with increasing density also many-body 
  correlations, in particular the appearance of clusters, i.e.\ 
  many-body bound states,
  are relevant. Presently  we include contributions 
  from $nn({}^{1}{S_{0}})$, $np({}^{1}{S_{0}})$,
  $pp({}^{1}{S_{0}})$ and $np({}^{3}{S_{1}})$ scattering channels and
  clusters with $A\leq  4$ in our calculations. 
  In the gRMF model light clusters (deuteron, triton, helion and
  $\alpha$-particle) 
  are introduced as additional degrees of freedom with temperature and 
  density dependent binding energy shifts given in Ref.\ \cite{Typ10}.
  Because we do not consider the formation of nuclei with mass numbers
  $A > 4$ the present model can only be applied to rather low densities
  where the fraction of heavier clusters can be neglected.

\subsection{Low densities}

  Similar as in the case of neutron matter, Figures \ref{fig:psym} and 
  \ref{fig:Esym} depict the quantities $p/n$ and $E/N$, calculated in different
  approaches, 
  for $T=4$~MeV and $T=10$~MeV as a function of the total particle
  number density $n$. 
  The extended gRMF model very well reproduces the VEoS at low densities, 
  deviating from it with increasing density and lower temperature due
  to medium effects. Note that in the depicted results of the
  VEoS calculation the same nuclei
  are considered as in the gRMF model. It is also worth noticing that the 
  standard RMF calculation without clusters and the extended gRMF
  model differ substantially.
  Again, we see the effect of the relativistic 
  corrections on the internal energy per baryon 
  $E/N$ in comparison with the ideal gas limit. 
  In contrast to the neutron matter case, where scattering 
  correlations are essential for reproducing the VEoS, in symmetric 
  matter the main contribution is caused by the appearance of
  bound states with positive binding energy.

\subsection{Composition}
\label{sec:snm_comp}

\begin{figure}[t]
  \begin{center}
    \epsfig{file=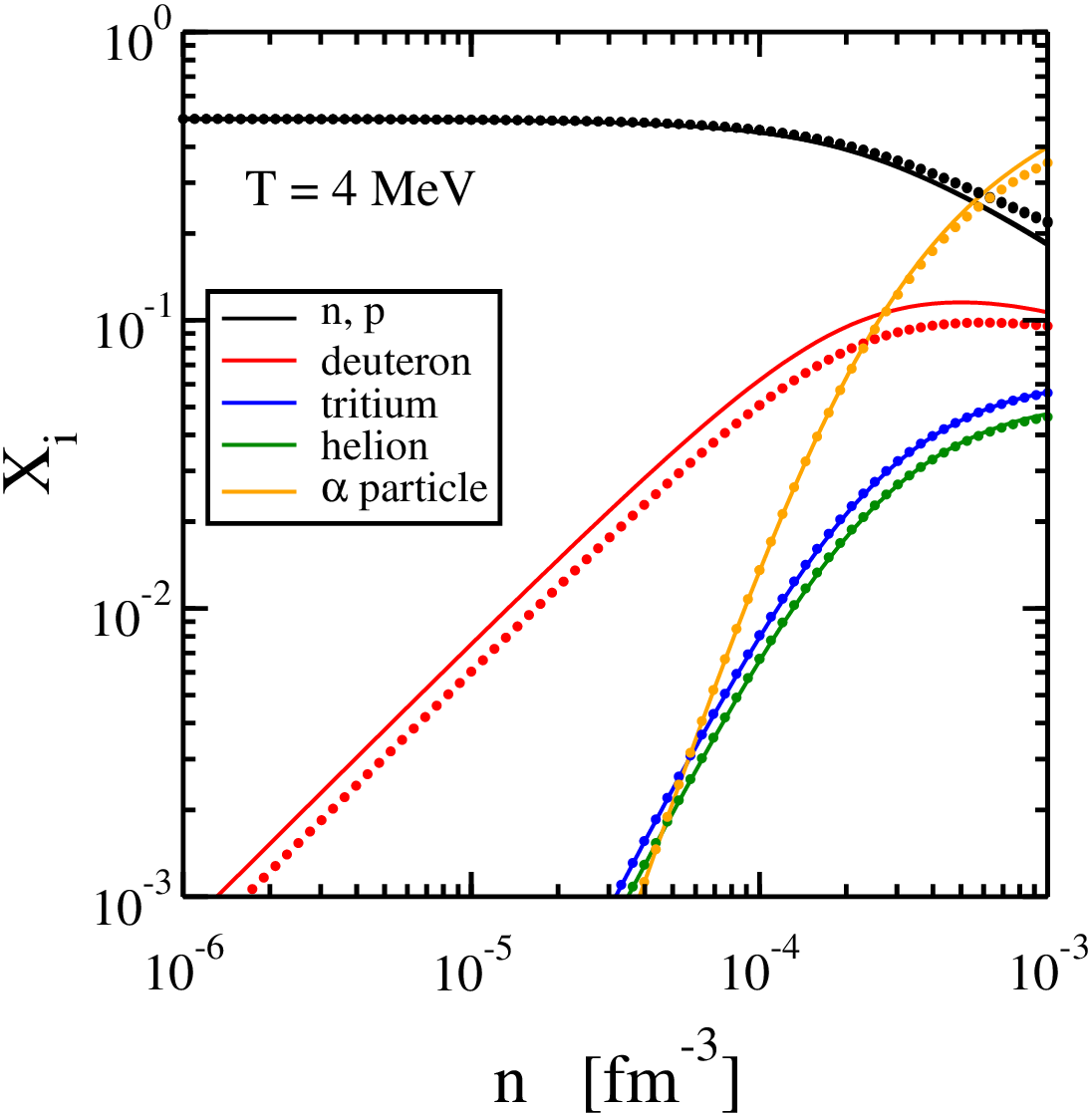,width=0.47\textwidth}
    \epsfig{file=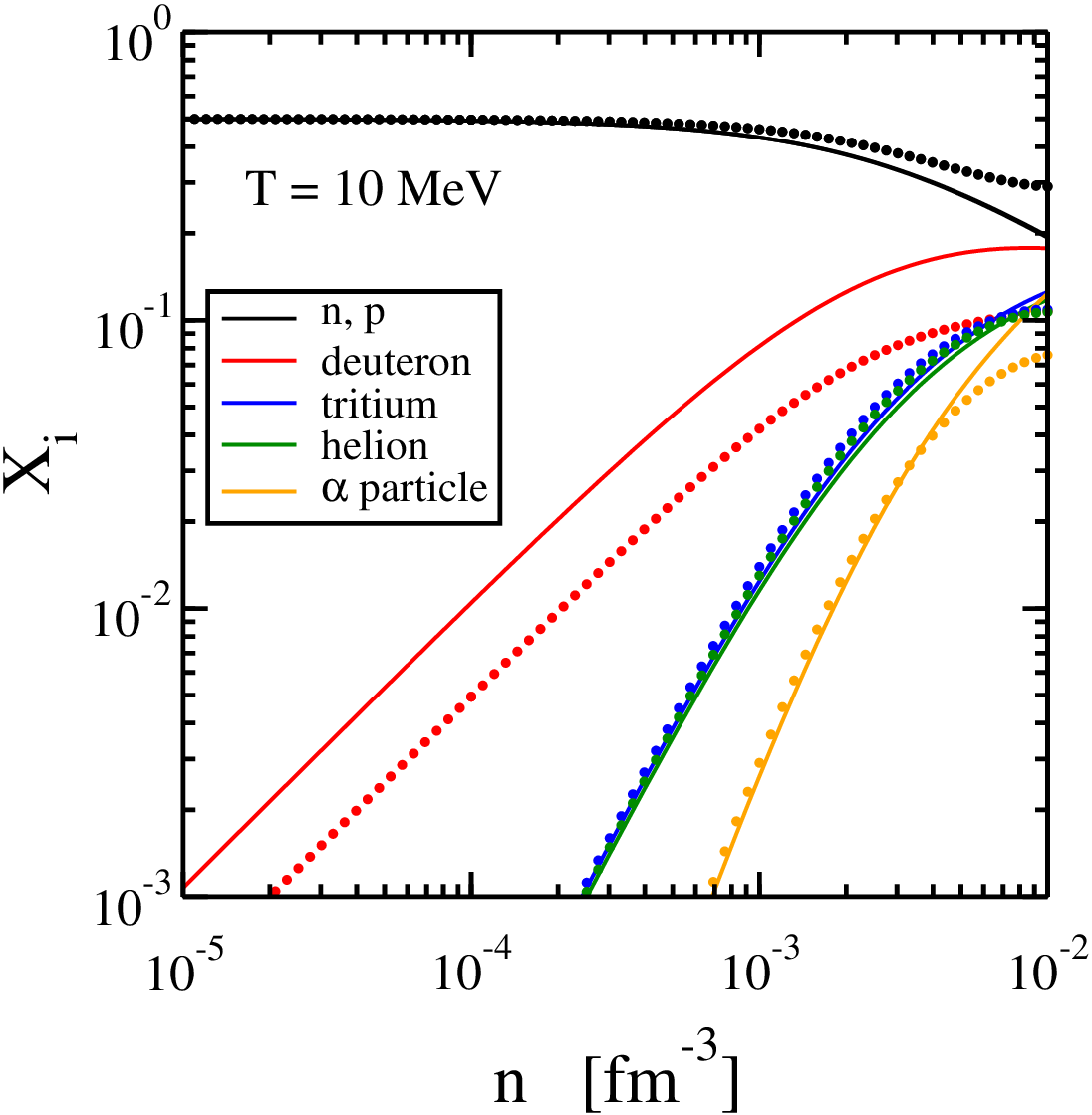,width=0.47\textwidth}
    \caption{\label{fig:Xsym}%
      Particle fractions $X_{i}$ of nucleons and light clusters
      in symmetric nuclear 
      matter as a function of the density $n$
      for temperatures of $T=4$~MeV (left) and $T=10$~MeV (right).
      Solid lines correspond to the VEoS and dotted lines to the
      gRMF model.}
  \end{center}
\end{figure}

It is worthwhile to study the detailed composition of symmetric
nuclear matter, 
comparing the fractions of various particles.
In Figure \ref{fig:Xsym} particle fractions $X_{i}=A_{i}n_{i}/n$ 
of nucleons and light clusters are shown for $T=4$ MeV and $T=10$ MeV for 
gRMF (dotted lines) and VEoS (solid lines) models.
The deuteron fraction $X_{d}$ contains the contribution of the
bound state and the isospin singlet $np$ scattering channel.
As is seen from the figures, in symmetric matter higher mass clusters 
become more important with increasing densities.
One notes that the particle fraction for deuterons in the VEoS and gRMF 
models are substantially different, especially in case for $T=10$
MeV. This behavior is caused by the fact that in the gRMF model part 
of the two-body correlation strength is shifted to the mean field, thus
reducing the contribution of the original two-body correlations. 
The same effect will be observed in case of neutron matter for the 
particular case of $nn$ correlations, see Section \ref{sec:high}. 
We also notice different 
slopes of the curves for light clusters, with the $\alpha$ particle having 
the steepest inclination. The reason is that at low total densities
cluster densities are 
proportional to fugacities in a certain power, e.g.\
 $n_d \sim z^2$,   $n_t \sim z^3$ and $n_{\alpha} \sim
 z^4$ with $z = z_{n} \approx z_{p}$ in symmetric nuclear matter. 
The overall scaling of the cluster fractions at low densities is
determined by the cluster binding energies. At densities $n$ above
$10^{-4}$~fm${}^{-3}$ and $10^{-3}$~fm${}^{-3}$ for $T=4$~MeV and
$T=10$~MeV, respectively, heavier nuclei will contribute significantly
to the composition in symmetric nuclear matter and they have to be
incorporated in the model calculations. Hence, it is not reasonable
to discuss the transition to higher densities in the present description
of symmetric nuclear matter. The case of pure neutron matter is
presented in the next section.

\section{Higher densities}
\label{sec:high}

In subsection \ref{subsec:nm_low} it was demonstrated that the gRMF approach
with effective $nn$ scattering correlations reproduces the low-density
limit of the VEoS for thermodynamical quantities such as $p/n$ and
$E/N$. With increasing density, the VEoS
approach is no longer applicable and a smooth transition of the gRMF
predictions to the
DD-RMF results with neutrons as quasiparticles is expected.
The details of this transition are affected by several ingredients of
the gRMF model that are not constrained by the low-density expansion:
the strength of the cluster-meson couplings (\ref{eq:Gim}),  the
shift $\Delta B_{nn}$ of the effective resonance energy $E_{nn}$ 
and possible contributions from three, four, \dots and
  many-neutron correlations.
In the following, only the variation of the transition with different
choices of $\Delta B_{nn}$ will be discussed. The strength of the
cluster-meson couplings is kept as given in the original gRMF model
and described in Section \ref{sec:gRMF}.

When the total density of neutron matter increases, the contribution
of the $nn$ cluster
in the gRMF model will be affected by the applied shift to the
resonance energy that represents the continuum correlations.
The functional dependence of the energy
shift $\Delta B_{i}$ of a cluster $i$ in (\ref{eq:Si}) with temperature $T$
and the meson fields $\omega$ and $\sigma$
is not determined by the low-density considerations.
Here, we explore three different choices. Similar as in
Ref.\ \cite{Typ10}, we write
\begin{equation}
 \Delta B_{i} = f[n^{\rm (eff)}_{i}] \delta B_{i}(T)
\end{equation}
with a function $f$ that depends on the effective density
\begin{equation}
 n^{\rm (eff)}_{i} = \frac{m_{\omega}^{2}}{\Gamma_{\omega}(0)} \omega
 + \frac{N_{i}-Z_{i}}{A_{i}} \frac{m_{\rho}^{2}}{\Gamma_{\rho}(0)} \rho
\end{equation}
and a temperature dependent factor $\delta B_{i}(T)$.
At low effective densities, $\Delta B_{i}$ should be linear in
$n^{\rm (eff)}_{i}$ and an obvious choice is $f = n^{\rm (eff)}_{i}$.
In Ref.\ \cite{Typ10} the quadratic form
\begin{equation}
 f=n^{\rm (eff)}_{i} \left[ 1+ \frac{1}{2}
 \frac{n^{\rm(eff)}_{i}}{n_{i}^{(0)}}\right]
\end{equation}
was used for the light clusters in order to obtain a stronger
suppression
of the cluster abundancies with increasing density. The density scale
$n_{i}^{(0)}= B_{i}^{\rm (vac)}/ \delta B_{i}(T)$ is set by the vacuum binding energy
$B_{i}^{\rm (vac)}$. Another possible choice is the pole form
\begin{equation}
 f = 
 \frac{n^{\rm (eff)}_{i}n_{\rm sat}}{n_{\rm sat}-n^{\rm(eff)}_{i}}
\end{equation}
for $n^{\rm(eff)}_{i}$ smaller than the saturation density 
$n_{\rm sat}$ of the gRMF model resulting in a complete dissolution of the
cluster when $n_{\rm sat}$ is approached from below because
$\lim_{n \to n_{\rm sat}} f = \infty$.

\begin{figure}[t]
  \begin{center}
    \epsfig{file=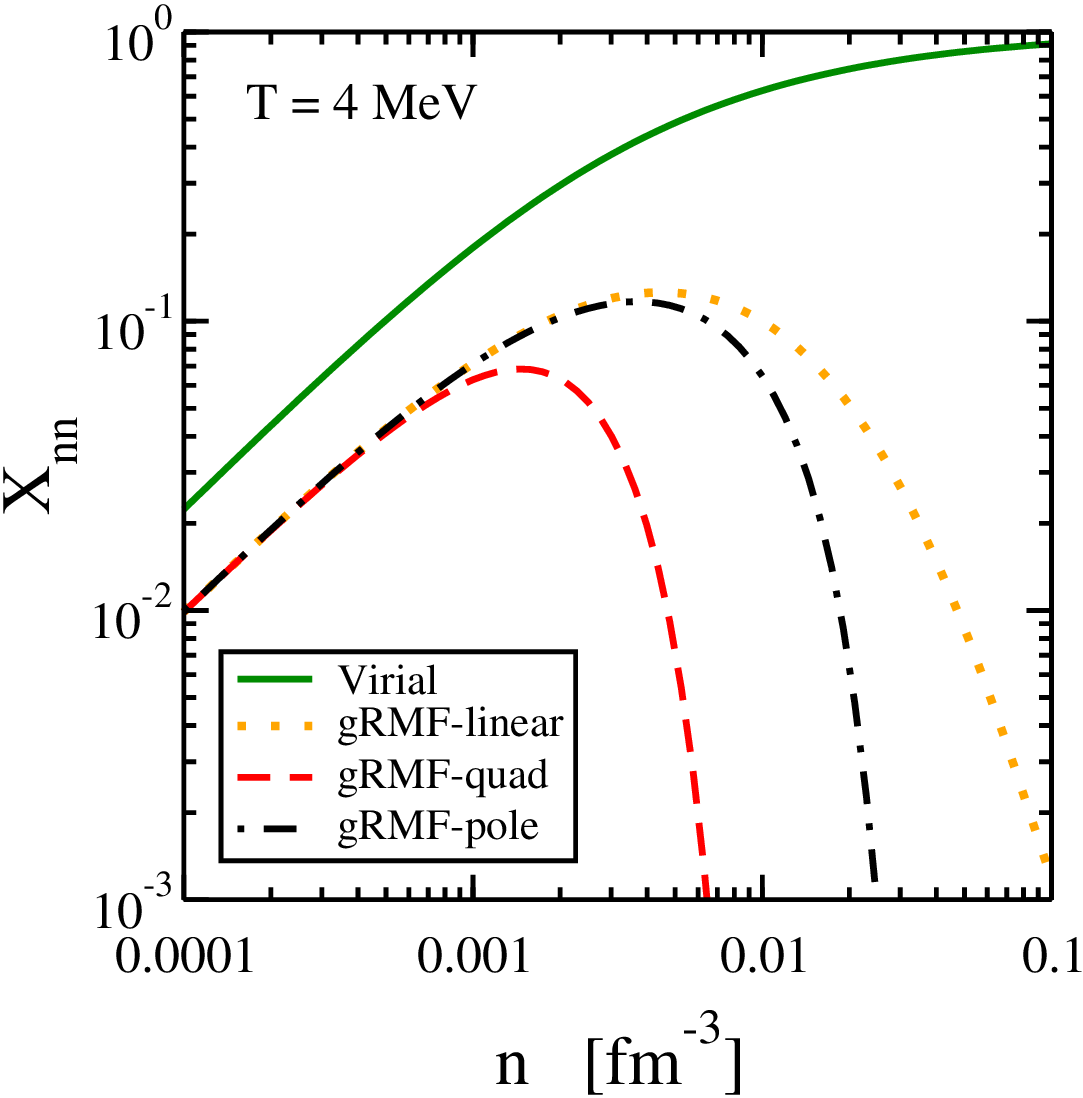,width=0.47\textwidth}
    \epsfig{file=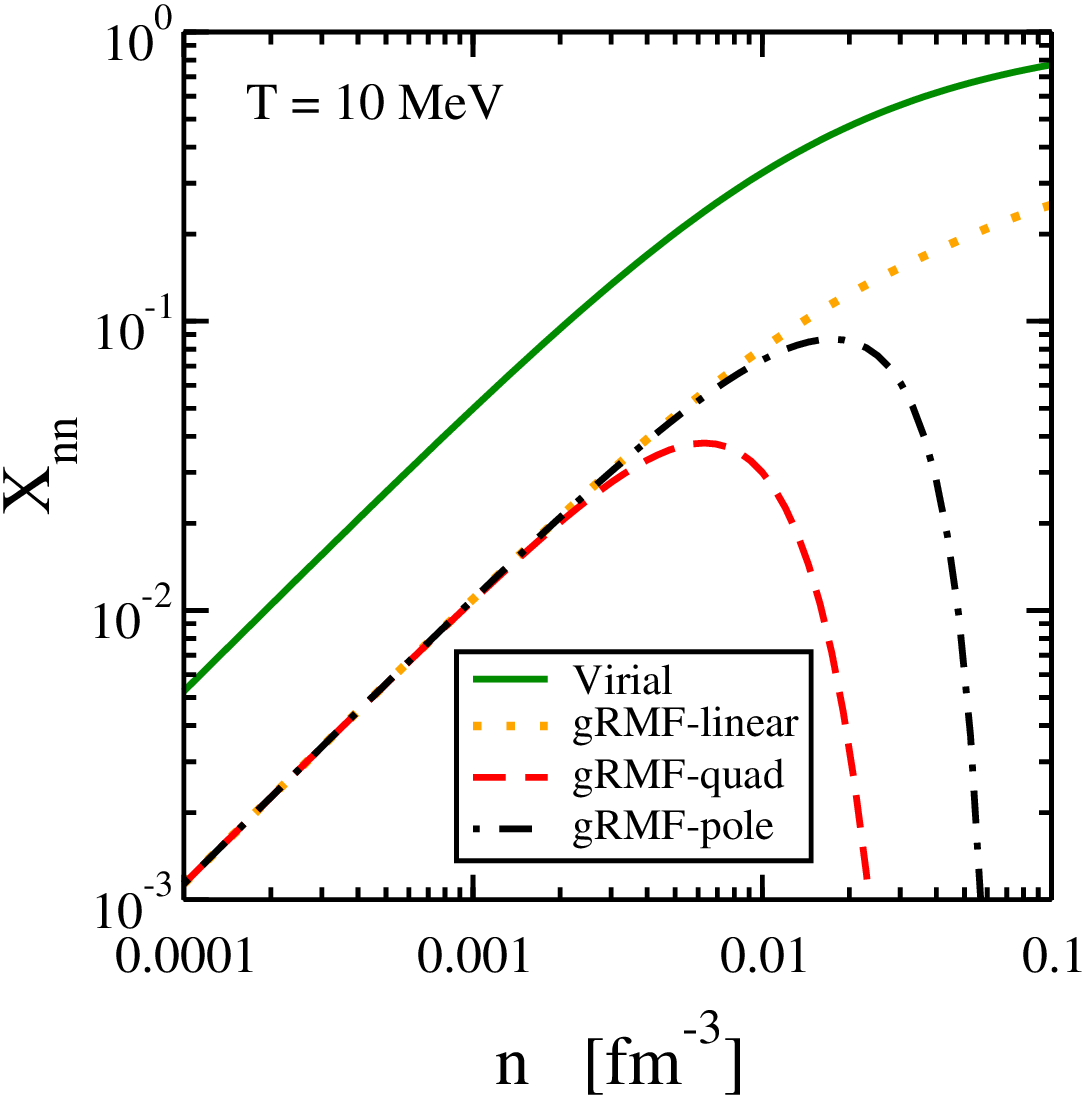,width=0.47\textwidth}
    \caption{\label{fig:Xnn}%
      Mass fraction $X_{nn}$ of the two-neutron correlation state
      in neutron matter as a function of the density $n$
      for temperatures of $T=4$~MeV (left) and $T=10$~MeV (right).}
  \end{center}
\end{figure}

\begin{figure}[t]
  \begin{center}
    \epsfig{file=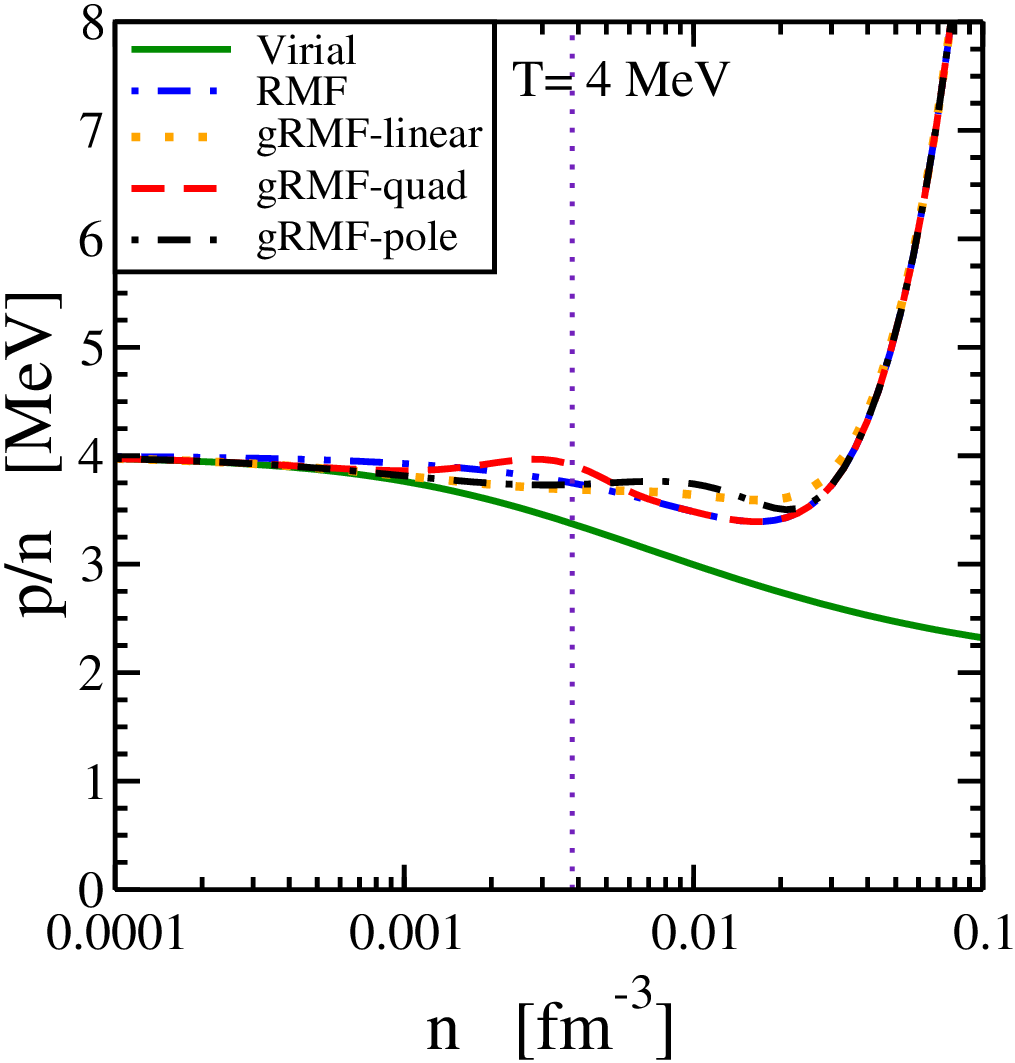,width=0.47\textwidth}
    \epsfig{file=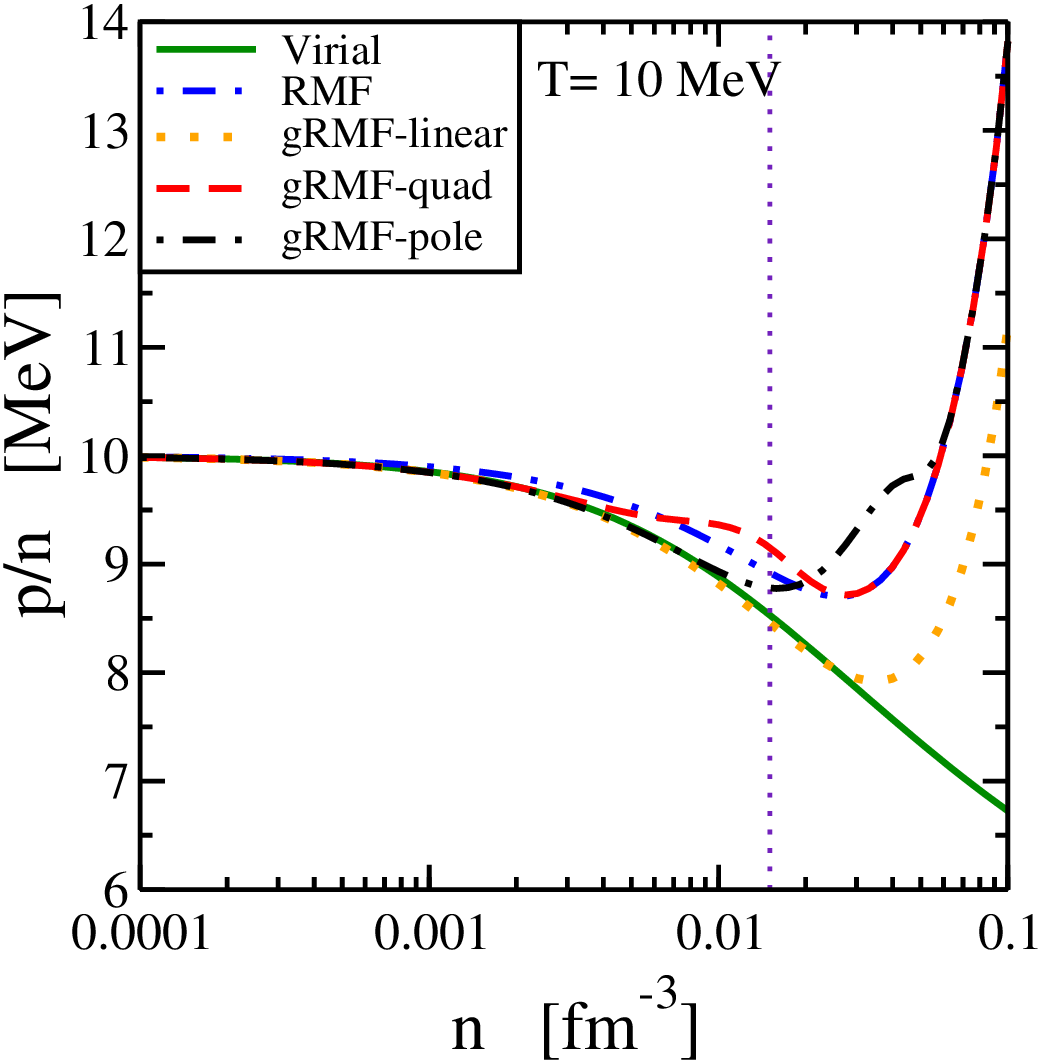,width=0.48\textwidth}
    \caption{\label{fig:pnhigh}%
      Ratio of pressure over total particle number density, $p/n$,
      of neutron matter as a function of the density $n$
      for temperatures of $T=4$~MeV (left) and $T=10$~MeV (right).
      Vertical dotted lines indicate the density where $n\lambda_{n}^{3}=1$.}
  \end{center}
\end{figure}

\begin{figure}[t]
  \begin{center}
    \epsfig{file=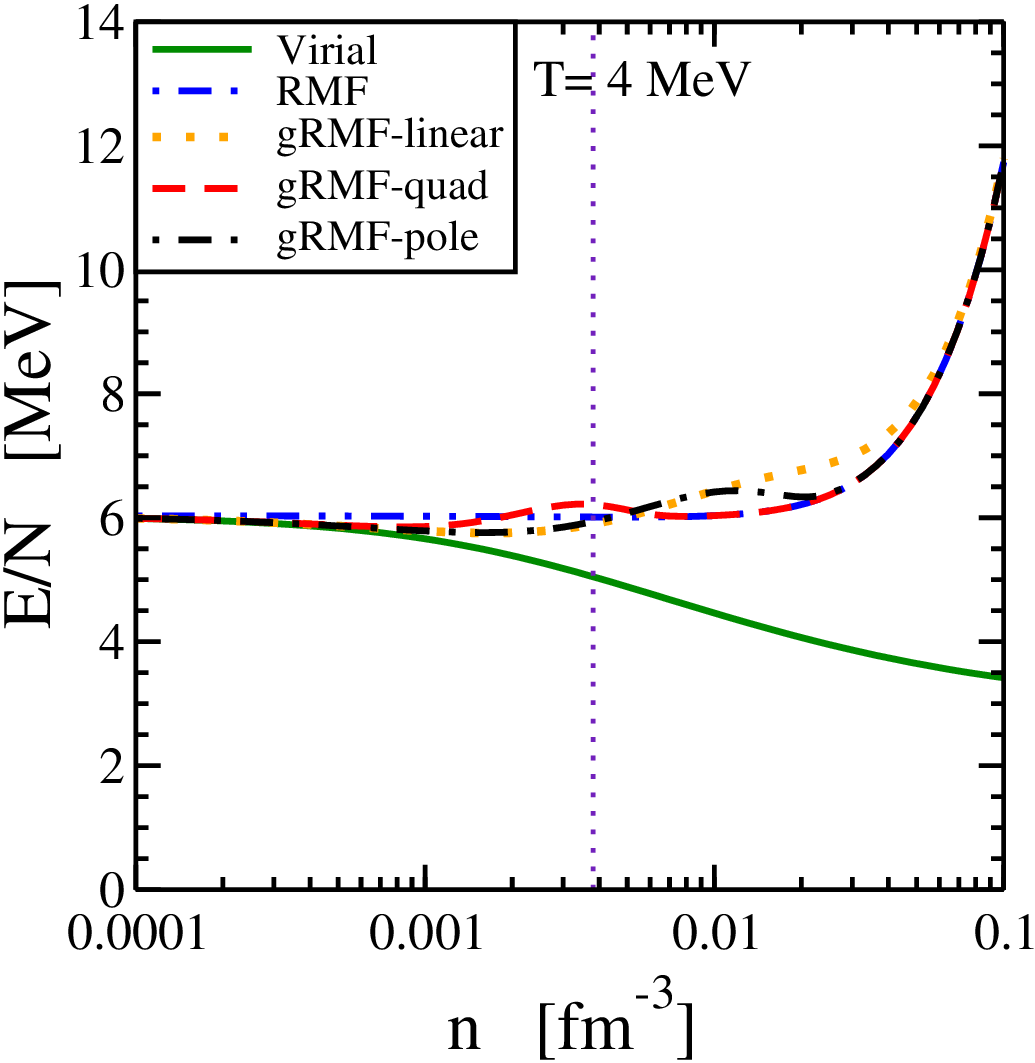,width=0.47\textwidth}
    \epsfig{file=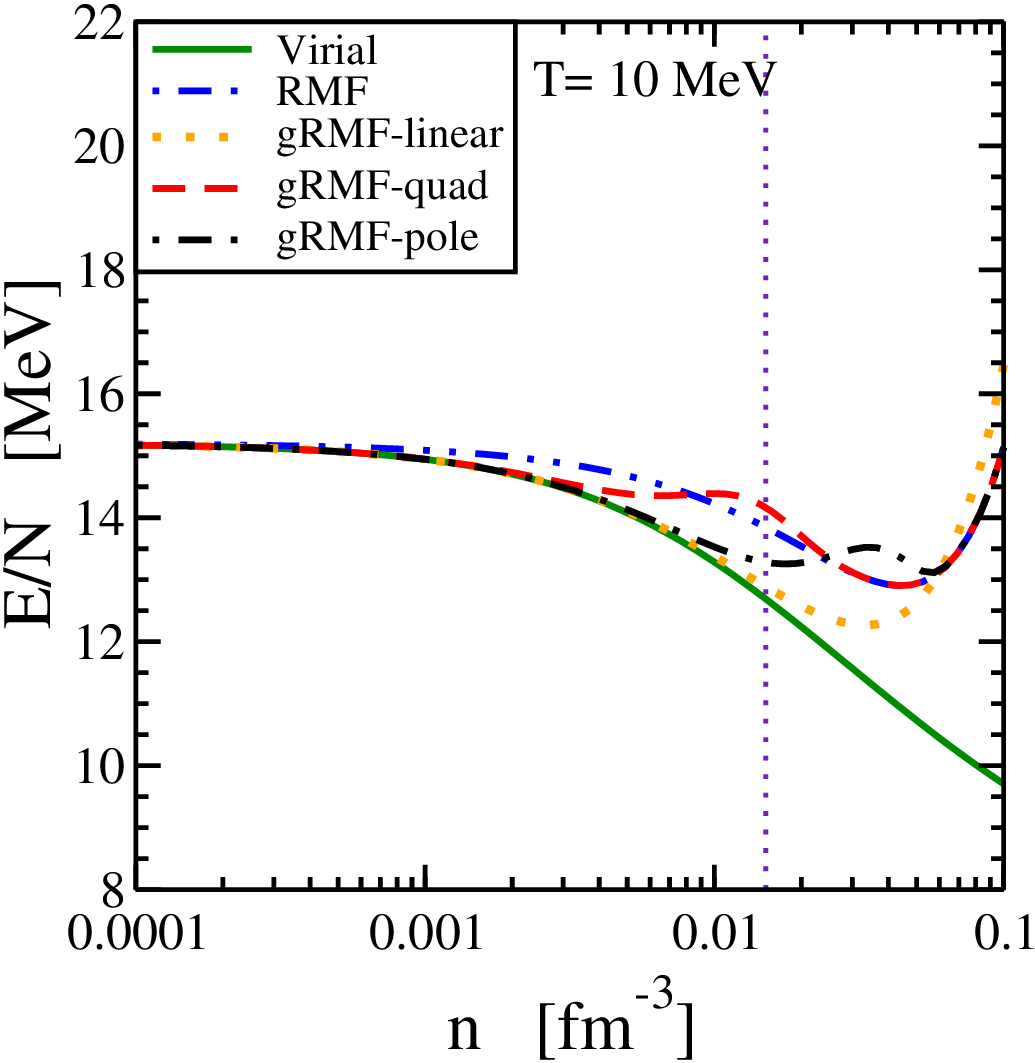,width=0.47\textwidth}\
    \caption{\label{fig:ENhigh}%
      Internal energy per baryon (without contribution of the neutron rest mass)
      in neutron matter as a function of the total particle number
      density $n$ for temperatures  $T=4$~MeV (left) and $T=10$~MeV
      (right). Vertical dotted lines indicate the density where $n\lambda_{n}^{3}=1$.}
  \end{center}
\end{figure}

The evolution of the two-neutron mass fraction
\begin{equation}
X_{nn} = 2n_{nn}/(n_{n}+2n_{nn})
\end{equation}
with increasing total density is depicted in Figure \ref{fig:Xnn}
for the VEoS and the gRMF model with the linear, quadratic and pole
form of the energy shift $\Delta B_{nn}$, respectively. We assume
$\delta B_{nn}(T) =\delta B_{d}(T)$ and set the density
scale $n_{nn}^{(0)} = n_{d}^{(0)} = B_{d}^{\rm (vac)}/ \delta B_{d}(T)$ with the
deuteron values for the quadratic dependence. The $nn$ fraction in the VEoS model rises
monotonously with the total density reaching unrealistically high
values much beyond the range of applicability of the approach.
At low densities, the $nn$ fractions for the different choices of the
energy shift in the gRMF model agree perfectly with each other.
They exhibit the same slope as the VEoS result.
In general, the $nn$ mass fraction at low densities is larger for the lower
temperature but the two-neutron cluster dissolves earlier with
increasing total density. The maximum mass fraction and range of
cluster dissolution depends sensitively on the form of the energy
shift $\Delta B_{nn}$. There are substantial variations that need to be
constrained in future investigations.
On an absolute scale, the gRMF
predictions for $X_{nn}$ are substantially smaller than those of the
VEoS at low densities. 
This difference is caused by the fact that in the gRMF approach
the correlations of quasiparticles are considered and part of the
correlation strength is contained in the self-energies, cf.\
the generalized Beth-Uhlenbeck approach in Subsection \ref{sec:GBU}.
The distribution of correlations between the explicit contribution
from the cluster state and the implicit contribution via the
self-energies depends on the nucleon-meson couplings at zero density
of the particular gRMF parametrization.

The dependence of the quantities $p/n$ and $E/N$ in neutron matter
on the total density is shown in Figures \ref{fig:pnhigh} and \ref{fig:ENhigh}
for temperatures $T=4$~MeV and $T=10$~MeV and a wider
range of densities compared to that shown 
in Figures \ref{fig:pn} and \ref{fig:EN}.  The vertical lines in Figures \ref{fig:pnhigh}
and \ref{fig:ENhigh} denote the density $n$ where $n \lambda_{n}^{3} = 1$. In the
low-density limit, all gRMF calculations reproduce the VEoS
predictions by construction but deviate from the DD-RMF model that does
not take cluster formation into account. At higher densities, the VEoS
fails to predict the strong increase of the pressure and
energy per neutron caused by the short-range repulsive $nn$ interaction.
The transition of the gRMF
results at low densities to the DD-RMF curve at higher densities
substantially depends on the choice of the energy shift $\Delta B_{nn}$ for the
effective resonance energy $E_{nn}$. A distinctive bump in $p/n$ and $E/N$
appears that is correlated with the sudden dissolution of the two-body
clusters
as depicted in Figure \ref{fig:Xnn}. This feature was already observed in
Ref.\ \cite{Typ10} for symmetric nuclear matter including only the
bound states of light clusters. The origin of this structure is
related to the contribution with the
derivatives of the energy shifts
in the $\omega$ and $\rho$-meson
field equations (\ref{eq:feq_o}) and (\ref{eq:feq_r}). 
The functional form
of the density dependence and absolute scale of the meson-cluster
coupling strengths  $\Gamma_{im}$ from (\ref{eq:Gim})
and (\ref{eq:gim}) will also have an impact on the
detailed form of the transition from the low-density limit to
higher densities. In this work, only the most simple choice of the 
factor $g_{im}$, proportional
to the number of nucleons in the cluster, was examined. Further
investigations are needed to fix the cluster-meson couplings and the
energy shifts less ambiguously.
Correlations beyond two-neutron states, that are not considered
for neutron matter in the present approach, could also modify
the features in the quantities $p/n$ and $E/N$.

\section{Conclusions}
\label{sec:concl}

In this paper, an extension of the generalized relativistic mean-field model with density
dependent couplings was developed by requiring the consistency of the
finite-temperature equation of state at low densities with the virial
equation of state that is a model independent benchmark in this limit.
For this purpose, new degrees of freedom were introduced in the 
generalized RMF
approach that represent two-nucleon correlations in the
continuum. These clusters can be considered as quasiparticles
like bound nuclei.
They are characterized by medium-dependent effective
resonance energies with temperature dependent effective degeneracy
factors.

From the comparison of the fugacity expansions of both the
relativistic mean-field and virial equation of state
models, consistency relations were derived that contain
the nucleon-nucleon scattering
phase shifts, the relativistic mean-field nucleon-meson coupling constants, the resonance
energies and effective degeneracy factors of the clusters. Various limits
were investigated for the choice of the relevant parameter functions and their
dependence on the thermodynamical quantities. For a successful
application of the approach, the density dependence of the 
original relativistic mean field model was kept and not modified at low
densities. The effective resonance energies were taken as given by
the calculation with the scattering phase shifts and the effective
degeneracy factors were assumed to be temperature dependent similar
as in the case of the nuclear statistical equilibrium model with 
thermally excited nuclei.

The example of neutron matter was studied in
detail for different temperatures. Relativistic effects were found to
become important with increasing temperature even at very low
densities. The extended generalized relativistic mean-field model 
smoothly interpolates between the
correct low- and high-density limits describing the dissolution of the
clusters. However, the precise form of the transition depends on the
coupling strength of the clusters to the meson fields and the energy shift
of the resonance energies. These quantities are not
fixed by the low-density constraints and the consequences of
different choices were investigated. This point deserves more studies
in the future. In the case of symmetric nuclear matter the
  formation of many-body bound states is crucial for a realistic
  description of correlations.

The proposed extension of the generalized relativistic {\rm mean-field} model
is rather general and can also be applied to other mean-field
approaches that aim at describing the equation of state of nuclear
matter. In the present work, only two-particle correlations were
considered. Bound states of light clusters with mass numbers $A \leq
4$ can be readily included as in Ref.\ \cite{Typ10}. 
A further extension to heavier nuclei with
medium-dependent binding energies and finite-temperature excitations
is also possible and corresponding work is in progress. 
For astrophysical applications, further contributions to the EoS than those
considered in this work have to be included in the model.
E.g.\ electrons, muons, photons, mesons etc. need to be considered
which also change the behavior of various thermodynamical quantities in
the low-density limit. Finally,
equation of state tables with the thermodynamical properties and the composition of
nuclear matter for a broad range in temperature, density and
proton-neutron asymmetry can be generated for 
astrophysical simulations of, e.g., core-collapse supernovae.

\section*{Acknowledgments}

The authors are grateful to D.\ Blaschke,
T.\ Fischer, M.\ Hempel, T.\ Kl\"{a}hn, G.\ R\"{o}pke, A.\ Schwenk, K.\ Vantournhout
and H.H.\ Wolter for discussions and comments during various stages of
the work. We thank M.\ Oertel for providing the data of the
Lattimer-Swesty EoS. The authors thank K.\ Langanke and
B.\ Friman for continuous interest in this work.
This research was supported by the DFG cluster of excellence `Origin
and Structure of the Universe' and by CompStar, a Research
Networking Programme of the European Science Foundation (ESF).
M.V.\ acknowledges support by the Helmholtz Graduate School for Hadron and
Ion Research (HGS-HIRe).
S.T.\ received support from
the Helmholtz International Center for FAIR (HIC for FAIR)
within the framework of the LOEWE program launched by the state of
Hesse via the Technical University Darmstadt and from the Helmholtz Association (HGF) through the Nuclear
Astrophysics Virtual Institute (VH-VI-417).




\appendix

\section{Virial equation in our model and in Ref.\ \protect\cite{Hor06b}}
\label{sec:HS}

The authors of Ref.\ \cite{Hor06b} use a slightly different
definition of some quantities in their approach to the non-relativistic 
VEoS compared to the ones we use in Section \ref{sec:veos}.
In order
to facilitate the comparison, we indicate the correspondence
of the two formulations.
Quantities of Ref.\ \cite{Hor06b} are indicated by a
$\check{\hspace{2mm}}$ in the following.

The single-particle partitions functions are
defined in the same way,
i.e.\ $Q_{i}=\check{Q}_{i}$ for nucleons and $\alpha$-particles,
however, there is a small difference in the thermal wavelengths because
the neutron and proton mass are assumed to be equal 
$\check{m}=\check{m}_{n}=\check{m}_{p}$
in Ref.\ \cite{Hor06b} and the $\alpha$-particle mass is set to
$\check{m}_{\alpha}=4\check{m}$ without considering the binding
energy as in Eq.\ (\ref{eq:m_i}). Similarly, for the non-relativistic
chemical potentials the relation
$\check{\mu}_{\alpha} = 2\check{\mu}_{n}+2\check{\mu}_{p}$
is used instead of
$\mu_{\alpha} = 2\mu_{n}+2\mu_{p}-B_{\alpha}$ in our case.
Nevertheless, the fugacities are identical. The main differences stem
from the definition of the many-body partition functions. In Eq.\
(\ref{eq:Qdef}) we place factors $1/n!$ in front of the $n$-body terms.
Consequently $\check{Q}_{ij} = Q_{ij}/2$.
Comparing Eq.\ (\ref{eq:Omega_series}) with Eq.\ (11) in \cite{Hor06b},
we identify
\begin{equation}
\label{eq:bnp}
 \check{b}_{n} = b_{nn}/2 = b_{pp}/2 ,\qquad
 \check{b}_{pn} = b_{pn}/2
\end{equation}
and
\begin{equation}
\label{eq:balpha}
 \check{b}_{\alpha} = b_{\alpha\alpha} , \qquad
 \check{b}_{\alpha n} = b_{\alpha n}/\sqrt{8} = b_{\alpha p}/\sqrt{8} \: .
\end{equation}
In Eq.\ (\ref{eq:vircoeff}) we use the c.m.\ energy $E$ as the
integration variable. In contrast, in Ref.\ \cite{Hor06b} the
laboratory energies $\check{E} = 2E$ are used in Eqs.\ (19), (22) and
(24) and the integrals are transformed with the help of a partial
integration with respect to $\check{E}$. Noting that
$\check{b}_{pn} = \check{b}_{\rm nuc} - \check{b}_{n}$ in Ref.\ \cite{Hor06b},
the formulas given there for $\check{b}_{n}$, $\check{b}_{pn}$ and $\check{b}_{\alpha}$
are consistent with the relations (\ref{eq:bnp}) and
(\ref{eq:balpha}). For the virial coefficient $\check{b}_{\alpha n}$, the
authors of Ref.\ \cite{Hor06b} use the nucleon laboratory energy
$\check{E}= 5E/4$ as integration variable. The expression
(26) in Ref.\ \cite{Hor06b}, however, is a factor two too large to be consistent
with the relation given in (\ref{eq:balpha}). This discrepancy was
already noted in Ref.\ \cite{Mal08}.

\section{Zero temperature low-density limit in the gRMF model}
\label{sec:RMFlow}

In the case of pure neutron matter
at zero temperature, all relevant thermodynamical quantities can be
represented analytically as a function of the Fermi momentum $k_{F_n}$.
The energy density $\varepsilon$ of pure
neutron matter without contributions of the rest mass
reads, cf., e.g., Ref.\ \cite{Typ99},
\begin{eqnarray}
\label{eq:eps}
 \varepsilon & = & \frac{3}{4} \sqrt{k_{F_n}^{2} + (m_{n}-S_{n})^{2}} \: n_{n}
 + \frac{1}{4} (m_{n}-S_{n}) \: n_{n}^{(s)} 
 \\ \nonumber & &
 + \frac{1}{2} \left[ \frac{\Gamma_{\omega}^{2}(n_{n})}{m_{\omega}^{2}}
 + \frac{\Gamma_{\rho}^{2}(n_{n})}{m_{\rho}^{2}} \right] n_{n}^{2}
 + \frac{1}{2} \left[ \frac{\Gamma_{\sigma}^{2}(n_{n})}{m_{\sigma}^{2}}
 + \frac{\Gamma_{\delta}^{2}(n_{n})}{m_{\delta}^{2}} \right] [n_{n}^{(s)}]^{2}
\end{eqnarray}
where the scalar neutron density is
given by
\begin{equation}
\label{eq:ns}
 n_{n}^{(s)} = \frac{3}{2x^{3}} f(x) \: n_{n}
\end{equation}
with the function
\begin{equation}
\label{eq:x1}
 f(x) = x\sqrt{1+x^{2}} - \ln  \left( x+\sqrt{1+x^{2}} \right)
\end{equation}
that depends on the dimensionless parameter $x=k_{F_n}/(m_{n}-S_{n})$.
We define the derivative of the function (\ref{eq:x1}),
 $f^{\prime} = 2x^{2}/\sqrt{1+x^{2}}$, and use the expansion
\begin{eqnarray}
(1+z)^{\alpha} & = & \sum_{k=0}^{\infty}
 \frac{\Gamma(\alpha+1)}{k!\Gamma(\alpha+1-k)} z^{2k}
 \end{eqnarray}
to rewrite equation (\ref{eq:ns}) after integration as
\begin{eqnarray}
\label{eq:nns}
 \frac{n_{n}^{(s)}}{n_{n}} &
  = &
 1 - \frac{3}{10} \left[\frac{k_{F_n}}{m_{n}-S_{n}}\right]^{2}
 +  \frac{9}{56} \left[\frac{k_{F_n}}{m_{n}-S_{n}}\right]^{4}
 + \dots
\end{eqnarray}
Substituting $n_{n}^{(s)}$  in (\ref{eq:eps}) 
by (\ref{eq:nns}) and doing subsequent expansions in powers of $k_{F_n}$ we arrive
at expression (\ref{eq:EN}) for the energy per neutron.

%

%
%
%






\end{document}